\documentclass[12pt]{article}
\pdfoutput=1
\usepackage{amsmath}
\usepackage{amssymb}
\usepackage{amsthm}
\usepackage{amscd}
\usepackage{cases}
\usepackage{amsfonts}
\usepackage{cite}
\usepackage{graphicx}
\usepackage{hyperref}
\usepackage[left=1in,right=1in,top=1in,bottom=1in]{geometry}
\DeclareGraphicsExtensions{.eps}
\usepackage{epstopdf}
\usepackage{xcolor}
\usepackage{fancyhdr}
\usepackage{lscape}
\usepackage{rotating}
\usepackage{bm}
\usepackage{ulem}
\usepackage{cancel}
\usepackage{tikz}
\usepackage{authblk}
\usetikzlibrary{arrows,backgrounds,decorations.pathreplacing}

\theoremstyle{plain} \numberwithin{equation}{section}
\newtheorem{theorem}{Theorem}[section]

\newtheorem{proposition}[theorem]{Proposition}
\theoremstyle{definition}

\usepackage{color}

\definecolor{darkgreen}{rgb}{0.1,0.4,0}

\begin{document}

\title{Temperature sensitivity of pest reproductive numbers in age-structured PDE models, with a focus on the invasive spotted lanternfly}

\author[1,*]{Stephanie M. Lewkiewicz}
\author[2]{Sebastiano De Bona}
\author[2]{Matthew R. Helmus}
\author[1]{Benjamin Seibold}
\affil[1]{Dept.\ of Mathematics, Temple University, Philadelphia, PA, USA}
\affil[2]{Dept.\ of Biology, Temple University, Philadelphia, PA, USA}
\affil[*]{Corresponding author (lewkiewicz@temple.edu)}
\maketitle
\begin{abstract}
Invasive pest establishment is a pervasive threat to global ecosystems, agriculture, and public health.  The recent establishment of the invasive spotted lanternfly in the northeastern United States has proven devastating to farms and vineyards, necessitating urgent development of population dynamical models and effective control practices.  In this paper, we propose a stage- and age-structured system of PDEs to model insect pest populations, in which underlying dynamics are dictated by ambient temperature through rates of development, fecundity, and mortality.  The model incorporates diapause and non-diapause pathways, and is calibrated to experimental and field data on the spotted lanternfly.  We develop a novel moving mesh method for capturing age-advection accurately, even for coarse discretization parameters.  We define a one-year reproductive number ($R_0$) from the spectrum of a one-year solution operator, and study its sensitivity to variations in the mean and amplitude of the annual temperature profile.  We quantify assumptions sufficient to give rise to the low-rank structure of the solution operator characteristic of part of the parameter domain.  We discuss establishment potential as it results from the pairing of a favorable $R_0$ value and transient population survival, and address implications for pest control strategies.
\end{abstract}

\textbf{Keywords}: age-structured PDE, stage-structured system, reproductive number, spotted lanternfly, pest population, moving mesh method\\

\textbf{2020 Mathematics Subject Classifications}: 35Q92 (Primary), 92-10, 92D40 (Secondary)


\section{Introduction}

Plant pests regularly impact ecological, human, and economic systems by catalyzing a cascade of structural ecosystem changes that result in biodiversity loss, alterations to nutrient cycles, increased air pollution, and other adverse effects\cite{LOVETT2016,CANELLES2021}. Economic globalization and climate change are amplifying these effects---the former through high rates of international trade that enable frequent introductions of species to nonnative regions, the latter through temperature conditions increasingly favorable for diverse species establishment across the globe~\cite{PURESWARAN2018}.

An emerging invasive insect in the United States is the spotted lanternfly (\textit{Lycorma delicatula}, SLF), a planthopper first detected in Berks County, Pennsylvania in 2014 after introduction via international shipping routes from China~\cite{URBAN2019,BARRINGER2015}. In Pennsylvania, this pest’s simultaneously high capacities for growth, spread, and impact on wild plants and agricultural products underscore the extent of the threat it poses~\cite{LEE2019}. Lanternfly presence in Pennsylvania vineyards has led to extensive loss in grape production, vine death, and lower fruit quality and yield, resulting in significant economic losses~\cite{URBAN2019}. In total, SLF is known to feed on at least 103 plant species worldwide, including 56 in North America, among them apple, walnut, willow, and maple~\cite{URBAN2019,BARRINGER2020}; however, its preferred host---on which it develops quickly---is the invasive tree of heaven (\textit{Ailanthus altissima})~\cite{DARA2015}, a weedy tree species native to China and Taiwan that grows in fragmented habitats worldwide~\cite{SLADONJA2015}. Adult lanternflies can fly for over a kilometer~\cite{WOLFIN2019}, but risk of long-distance spread is highest for eggs because gravid females lay eggs on a plethora of both natural and man-made surfaces, including flat metal, like that of cargo containers and train cars, and wood, like that in shipping pallets~\cite{URBAN2019}. In the last half-decade, the core population has expanded its range from Pennsylvania to New Jersey, New York, Connecticut, Massachusetts, Virginia, Delaware, Maryland, Ohio, and Indiana~\cite{LEE2019,NYSIPM2021}. This current invasion---while particularly dangerous---is merely one in a long-term pattern of nonnative insect plant pest introductions.

In this paper, we model the population dynamics of SLF in fixed locations using parameters obtained from experimental data on SLF development, mortality, and reproduction. However, our modeling approach is general and may be applied to a plethora of pests with dynamics dictated by time-dependent climatic conditions, such as seasonal variation in temperature.  The specific question that our model aims to answer is: how suitable is the seasonal temperature environment in a given location for pest population establishment? To model a local insect population, we consider an age-structured, stage-structured, linear system of PDEs that captures, mechanistically, the effect of seasonal temperature variation on different pest life stages through rates of development, fecundity, and mortality. A linear model is appropriate for this problem due to the following assumptions: (a)~abundant host resources are available for the pest in the given location, (b)~the population is large enough that positive-density dependent effects, like the Allee effect, are negligible, and (c)~the population is small enough that negative-density dependent effects, like overpopulation or other bug--bug interactions, are negligible.  Such assumptions are especially valid for the nascent introductions with which we are primarily concerned, in which populations have been observed to grow at rapid rates following establishment.  

The sensitivity of population dynamics to temperature patterns has been explored thoroughly in the case of the mosquito~\cite{WANG2016,LAMBRECHTS2011}, as well as other destructive or disease-carrying pests~\cite{EWING2021,LIU2017}, using compartmental stage-structured ODE systems~\cite{LIU2002}.  The incorporation of continuous age-structures within the stages of a compartmental model allows for a more nuanced description of age-dependent dynamics.  Here, this step enables us to model the attenuation of reproductive capacity as individuals age within the egg-laying stage, and, in turn, capture the sensitivity of egg-laying rates to the temporal interactions between temperature patterns and distributions of individuals in the egg-laying stage.  Age-structured equations have long been used in biological modeling\cite{KEYFITZ1997,SHARPE1911,IANNELLI2017}, including numerous recent advances in insect population dynamics.  (For a thorough description of the generic single- and multi-stage age-structured system used for insects, see~\cite{BUFFONI2007}.)  Model behavior has been studied numerically~\cite{PASQUALI2019,IANNELLI2017,IANNELLI1997,KAKUMANI2018,HE2018} and analytically\cite{BUFFONI2007,IANNELLI2017,CUSHING1994,GURTIN1974,HE2018}, and models have been calibrated to numerous pests, including the grape berry moth~\cite{GILIOLI2015}, apple snail~\cite{GILIOLI2017}, and codling moth~\cite{PASQUALI2019}.  PDE model analysis and calibration has yet to be conducted on the spotted lanternfly, despite significant interest in this pest.  To this end, we formulate generic coefficient functions describing rates of development, fecundity, and mortality, and calibrate them based on emerging experimental data and insight.

Whereas previous modeling efforts have been designed around a singular pathway through the life cycle~\cite{BUFFONI2007}, we expand this framework to allow for multiple pathways through early development.  We apply this strategy to explicitly model diapause, a period of cessation of development adopted by many insect species to overcome adverse environmental conditions, as a distinct life stage~\cite{KOSTAL2006,GILL2017}. The timing of diapause, both from a seasonal and a developmental perspective, varies between species: in SLF and many other insects in temperate climates, diapause occurs at the egg stage to overcome harsh winter conditions.  In this case, the diapause process tends to synchronize the developmental ages of a population of eggs laid at different times~\cite{KEENA2021,TAUBER1976,SAUNDERS1981}, and its incorporation into an age-structured PDE model can manifest as solution functions (age distributions) with support of very small measure.  When numerical approximations to such functions are computed with standard finite volume methods, they tend to be deformed significantly by numerical (diffusion/dispersion) effects, resulting in a high degree of sensitivity to numerical parameters.  As a remedy, we develop a moving mesh method that avoids these numerical diffusive effects, leading to accurate computations of the solution with relatively coarse discretizations.

Employing this method, we compute the dominant eigenvalues of a one-year solution operator, regarding the largest eigenvalue as a proxy for the one-year reproductive number, $R_0$.  To broadly characterize the sensitivity of $R_0$ to different temperature patterns, we model the annual temperature profile with a sinusoidal function and conduct a  parameter sweep of the two dominant eigenvalues with respect to mean and amplitude.  Under the diapause model, a low rank structure manifests in the solution operator within the moderate-mean/moderate-amplitude region of the parameter domain.  We elucidate a set of temperature conditions and physiological assumptions sufficient to produce this low rank structure.  We then analyze the remaining qualitatively distinct regions of the parameter domain under diapause and non-diapause models, and study the transient behavior resulting from initial conditions representing realistic instances of population dispersal to new locations.  From the coupling of these predictions of transient and asymptotic solution behavior ($R_0$), we discuss implications for pest establishment in climatically diverse regions.  We conclude by plotting the range of the United States in the model's phase space, which provides clear predictors of SLF establishment potential in the US.

The remainder of the paper is organized as follows: In Sec.~\ref{sec:mathmodel}, we present the mathematical model and its calibration to real-world data. In Sec.~\ref{sec:numericalmethod}, we present the moving mesh method for a one-dimensional advection equation and then describe the full numerical method for solving the model equations. In Sec.~\ref{sec:reproductivenumber}, we define $R_0$, present the results of the parameter sweep, discuss the advantages of our numerical method in computing $R_0$ accurately, and present simulations illustrating transient behavior. Finally, in Sec.~\ref{sec:summary}, we summarize the insight into establishment potential provided by our model.

\section{Mathematical Model}
\label{sec:mathmodel}

\begin{figure}[htbp]
    \centering
    \tikzstyle{diap egg} = [fill=blue!20,draw,rectangle,double,rounded corners,align=center,below,inner sep=0pt,minimum height=16mm,minimum width=30mm]
    \tikzstyle{dev egg} = [fill=teal!20,draw,rectangle,double,rounded corners,align=center,below,inner sep=0pt,minimum height=16mm,minimum width=30mm]
    \begin{tikzpicture}[>=stealth,scale=0.9,transform shape]
    \path (-10,-1.2) node [diap egg] (diapauseeggs) {diapausing\\eggs ($d$)}
          (-6,-1.2) node [dev egg] (postdiapauseeggs) {post-diapause\\eggs ($p$)}
          (-8,1.2) node [dev egg] (nondiapauseeggs) {non-diapause\\eggs ($u$)}
          (-1,0.3) node[fill=violet!20,draw,rectangle,double,rounded corners,align=center,below,inner sep=0pt,minimum height = 24mm,minimum width=40mm] (instars) {instars}
          (2,0.3) node[fill=yellow!20,draw,rectangle,double,rounded corners,align=center,below,inner sep=0pt,minimum height = 24mm,minimum width=20mm] (youngadult) {young\\adults}
          (4,0.3) node[fill=red!20,draw,rectangle,double,rounded corners,align=center,below,inner sep=0pt,minimum height = 24mm,minimum width=20mm] (egglayers) {egg-layers};
          \draw [thick,decorate,decoration={brace,mirror,raise=5pt,amplitude=14pt}] (-3,-2) -- (5,-2) node[pos=0.5,below=20pt,black]{$(b)$};
          \draw[->,ultra thick,rounded corners] (egglayers.north) to ++(0,2cm) to ++(-120mm,0) to (nondiapauseeggs.north);
          \draw[->,ultra thick,rounded corners] (egglayers.south) to ++(0,-2cm) to ++(-140mm,0) to (diapauseeggs.south);
          \draw[->,ultra thick] (nondiapauseeggs.east) to [out=0,in=180] (instars.west);
          \draw[->,ultra thick] (postdiapauseeggs.east) to [out=0,in=180] (instars.west);
          \draw[->,ultra thick] (diapauseeggs.east) to (postdiapauseeggs.west);
          \draw[->,ultra thick] (-1,-0.15) to (4,-0.15);
          \path (-10,2.1) node [thick,draw,rectangle,rounded corners,fill=white!] {SESSILES};
          \path (-1.5,1.1) node [thick,draw,rectangle,rounded corners,fill=white!] {MOTILES};
          \path (-2.2,2.6) node {winter/spring};
          \path (-3,-4.5) node {summer/fall};
          \begin{pgfonlayer}{background}
                \fill [fill=black!10]
                (-11.75,-3.3) rectangle (-4.25,2.7);
                \fill [fill=black!10]
                (-3.3,-3.5) rectangle (5.25,1.7);
          \end{pgfonlayer}
    \end{tikzpicture}
    \caption{\textbf{Spotted Lanternfly life cycle.} Eggs are referred to as \textit{sessiles}, while juveniles (\textit{instars}) and adults are referred to as \textit{motiles}. Newly-hatched lanternflies pass through the instar, young adult, and egg-laying stages sequentially.  Eggs laid in the summer and fall are assumed to pass through diapause before beginning development post-diapause. Eggs laid in the winter and spring are assumed to bypass diapause, developing instead as non-diapause eggs. After hatching, all instars and adults are regarded as physiologically identical, regardless of whether or not they passed through diapause as eggs.}
    \label{fig:lifecycle}
\end{figure}
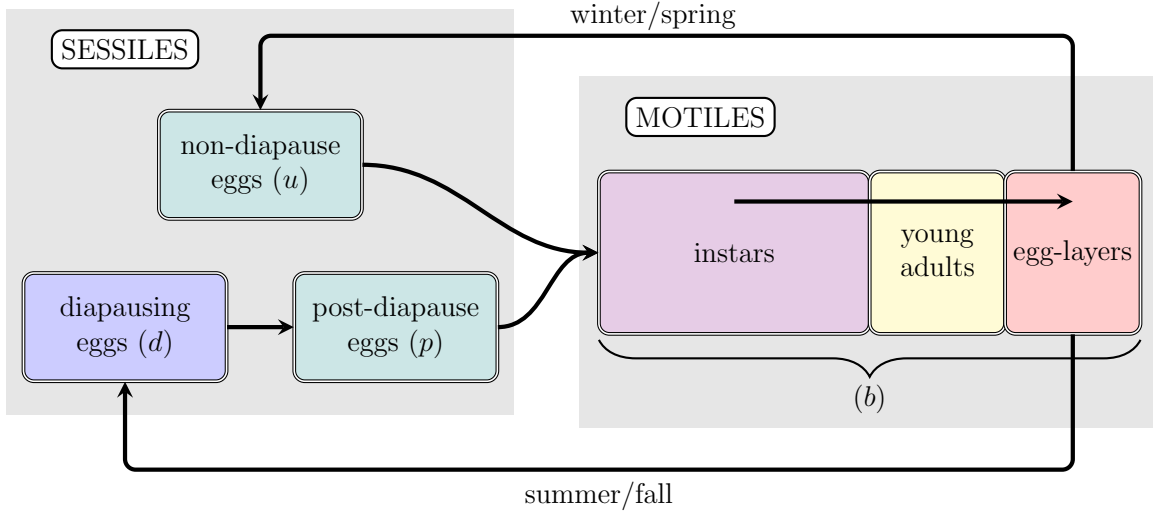

To model the dynamics of an insect population, we consider a stage-structured system of age-structured PDEs in four density functions corresponding to the quantitatively distinct life stages through which an individual may pass.  These four functions arise by subdividing the population into three sessile stages (diapausing eggs, eggs developing post-diapause, and eggs that are developing without having experienced diapause) and one motile stage (juveniles---often referred to as \textit{nymphs} or \textit{instars}---and adults).  Although individuals may present different physiological characteristics at different points in a given stage, individuals in the same stage are united in their common response to temperature, which drives rates of development and mortality.  While age-structured PDEs are not uncommon in mathematical population modeling efforts~\cite{IANNELLI2017}, the specific model formulation here differs fundamentally from other recent work in its incorporation of distinct diapause and non-diapause developmental pathways.  Diapause is a period of cessation of metabolic activity induced to protect eggs from extreme cold and synchronize a population developmentally~\cite{KEENA2021}.  Eggs are laid by reproductive-age adult females, and are assumed to either be laid in diapause or have committed to develop without undergoing diapause at the time of egg-laying, in accordance with diapause-induction criteria determined by the photoperiod experienced by a fecund female~\cite{SAUNDERS2004,MOUSSEAU1991}; in particular, we assume that the photoperiods of summer and fall trigger diapause, while those of winter and spring do not.  Upon completion of diapause, eggs resume normal development to the point of hatching.  Eggs in which diapause was not triggered develop normally from the moment of egg-laying, albeit without the lasting protective effects conferred on the egg by the diapause process, which are assumed to remain even after diapause termination~\cite{THOMAS2012}.  Newly hatched individuals pass through four instar stages before adult emergence and later reproductive maturity~\cite{DARA2015,LIU2019}, with all of these stages collectively comprising the motile group.  This basic life cycle is depicted in Fig.~\ref{fig:lifecycle}.

\subsection{Mathematical Structure}

To define our PDE system in the framework of the four-stage life cycle described above, we denote by $I$ the index set $I := \{u, d, p, b\}$, where $u$, $d$, and $p$ represent the non-diapause, diapause, and post-diapause sessile egg stages, respectively, and $b$ represents the motile stage.  The function $\rho^i(a,t)$ for $i \in I$ denotes the density (with respect to $a$) of female individuals in the stage associated with $i$ having developmental age $a$ at time $t$.  Here, $t$ is measured in days, while $a$ is a unitless variable that represents the fraction of the individual's current life stage through which it has advanced, so that $a \in [0,1]$.  As rates of development (resp., diapause advancement) in insects vary with the temperatures to which individuals are exposed (\textit{poikilothermy})~\cite{WAGNER1984}, the variable $a$ represents a developmental (resp., diapause) age distinct from chronological age. 

Denoting by $\Omega$ the domain $(0,1) \times (0,\infty)$, the functions $\rho^i(a,t)$ for $i \in I$ are given by solutions of the PDEs,
\begin{align}
\rho^i_t + \nu^i(t) \rho^i_a - \sigma^i \nu^i(t) \rho^i_{aa} + m^i(t) \rho^i = 0 \ \text{ for } (a,t) \in \Omega, ~\label{eq:genpde}
\end{align}
subject to the initial conditions,
\begin{align}
\rho^i(a,0) = \rho_0^i(a) \ \text{ for } a \in [0,1], ~\label{eq:initcond}
\end{align}
and the flux balance conditions,
\begin{align}\begin{split}
\alpha \nu^b(t) \int_{0}^1 \rho^b(a,t) k(a) \textrm{d}a =&\\
[\nu^u(t) \rho^u(0,t) - \sigma^u \nu^u(t) \rho_a^u(0,t)] \xi(t)& + \left[\nu^d(t) \rho^d(0,t) - \sigma^d \nu^d(t) \rho_a^d(0,t) \right](1-\xi(t)),\end{split} \label{eq:fluxbalance1}\\[1em]
\nu^p(t) \rho^p(0,t) - \sigma^p \nu^p(t) \rho^p_a(0,t) & =\nu^d(t)\rho^d(1,t) - \sigma^d \nu^d(t) \rho_a^d(1,t),\label{eq:fluxbalance2}\\[1em]
\left[\nu^b(t) \rho^b(0,t) - \sigma^b \nu^b(t) \rho^b_a(0,t)\right] &= \left[\nu^u(t) \rho^u(1,t) - \sigma^u \nu^u(t) \rho^u_a(1,t)\right] \nonumber\\
& + \left[\nu^p(t) \rho^p(1,t) - \sigma^p \nu^p(t) \rho^p(1,t)\right],\label{eq:fluxbalance3}
\end{align}
for $t \in (0,\infty)$.  In Eq.~\eqref{eq:genpde}, $\nu^i(t)$ (units: $\text{day}^{-1}$) denotes the rate of development for $i \in I/\{d\}$, and that of diapause advection for $i = d$.  Due to insect poikilothermy, these advective rates depends on $t$ through temperature, $T$ (units: $^o C$), so that $\nu^i(t) = \nu^i(T(t)) \geq 0$.  Development typically occurs when temperatures exceed some minimal threshold, $T_{\nu^i,1}$, with development rates increasing with temperature until a plateau is reached at high tempertures~\cite{SMYERS2021}.  We therefore restrict ourselves to nondecreasing, continuous functions $\nu^i(T)$ ($i \in I/\{d\}$) that satisfy: $\nu^i(T) = 0$ for $T < T_{\nu^i,1}$, and $\nu^i(T) \to \nu^{i,\textrm{max}} > 0$ as $T \to \infty$.  Diapausing eggs typically advance through diapause fastest within an optimal, cool temperature range, $(T_{\nu^d,1},T_{\nu^d,2})$, with slower advancement at more extreme temperatures~\cite{SHIM2015}.  In the case $i = d$, then, we consider continuous functions $\nu^d(T)$ satisfying $\nu^d(T) = \nu^{d,\textrm{max}}$ for $T \in (T_{\nu^d,1},T_{\nu^d,2})$, $\nu^d(T) \to \nu^{d,\textrm{min}}$ as $T \to \pm \infty$, where $\nu^{d,\textrm{max}}, \nu^{d,\textrm{min}} > 0$ are the absolute maximum and minimum values of $\nu^d$.  Here, we assume that $\nu^{d,\textrm{min}}>0$ to ensure that eggs do not become ``trapped" in diapause indefinitely.

The parameters $\sigma^i \geq 0$ (units: $\text{day}^{-1}$) in Eq.~\eqref{eq:genpde} are diffusion constants that account for variability in actual rates of development among individuals in each life stage.  For $i \in I/\{d\}$, $\sigma^i > 0$. We take $\sigma^d = 0$ for simplicity, due to a lack of knowledge of interindividual variation in the metabolic protein-clearing processes that drive rates of diapause advancement.  The functions $m^i(t)$ (units: $\text{day}^{-1}$) are per-capita death rates, which, in general, may account for basal mortality and/or mortality due to extreme temperatures, so that $m^i(t) = m^i(T(t)) \geq 0$.  The sessile mortality functions $m^i(T)$ for $i \in I/\{b\}$ account for temperature-induced mortality alone, and reflect the fact that temperatures below a certain threshold temperature, $T_{m^i,1}$, are more lethal, while those above the threshold do not induce excess death.  Thus, we consider continuous, nonincreasing functions $m^i(T)$ that satisfy $m^i(T) = 0$ for $T > T_{m^i,1}$ and $m^i(T) > 0$ for $T < T_{m^i,1}$.  We account for basal mortality in sessile individuals through the factor $\alpha \in [0,1]$ in Eq.~\eqref{eq:fluxbalance1}, which represents the fraction of the total egg population that would survive at temperatures known to not induce excess mortality in eggs.  In this case, $\alpha$ acts as a filter at the time of egg-laying, preliminarily culling all eggs destined to perish at any point in the sessile stage for reasons outside of exposure to harsh temperatures.  In contrast, the motile mortality function, $m^b(T)$, is chosen to account for both basal and temperature-induced death; the precise age distribution of late-stage motiles dictates reproduction rates, thus individuals must be accounted for in the system until the actual moment of death.  We therefore consider continuous functions $m^b(T)$ having the properties: $m^b(T) = m^{b,\textrm{min}}$ for $T \in (T_{m^b,1},T_{m^b,2})$, $m^b(T) \to \infty$ as $T \to \pm \infty$, where $m^{b,\textrm{min}}$ represents the basal death rate and $(T_{m^b,1},T_{m^b,2})$ is the ideal temperature range.

In Eq.~\eqref{eq:fluxbalance1}, the binary function $\xi(t)$ acts to direct newly-laid eggs into the diapause or non-diapause egg domains based on the assumption that diapause induction is a matter of seasonality.  While $t$ represents the number of days elapsed since the start of the simulation, each time $t$ is also associated with a specific date and time, so that diapause induction may depend on calendar date.  We take $\xi(t) = 1$ if the calendar date $t$ falls after the summer solstice and before the next winter solstice, and $\xi(t) = 0$ otherwise.  We note that the assumption that seasonality affects population dynamics through diapause induction relies, in turn, on an assumption that SLF can perceive day-to-day variation in photoperiod and infer coming climatic changes (e.g. temperature increases or decreases).  While it is known that SLF are generally capable of perceiving variations in photoperiod, the magnitude of the day-to-day variation required for them to detect the change is not known.  As such, we restrict our modeling efforts to regions in which photoperiods exhibit some annual variance (i.e., regions away from the equator).  Also in Eq.~\eqref{eq:fluxbalance1}, the egg-laying kernel, $k(a) \geq 0$, represents the density of female eggs laid by a female of age $a$.  We require that $k(a)$ be an integrable function that is non-increasing over its support, as an individual's capacity for laying eggs should decrease with age due to exhaustion of metabolic resources~\cite{FOX1993}.  Then $\int_0^1 k(a) \mbox{ }\textrm{d}a$ represents the total female eggs one female lays over the course of her life.

\subsection{Calibration of the Model Parameters and Functions}
\label{sec:calibration}
We now describe how the generic model functional forms and parameters described above can be tailored and calibrated to the biology of the spotted lanternfly.  The relative developmental age variable, $a$, is defined for each of the four life stages with respect to the duration of each stage in typical developmental units.  For all life stages of the system except diapausing eggs, the developmental unit is the spotted lanternfly \textit{degree day}---a measure of heat accumulation, which, in turn, tracks development~\cite{SMYERS2021}.  Developing individuals accumulate degree days throughout periods of exposure to sufficiently warm temperatures, and, once certain threshold degree day counts are reached, transition to the next life stage.  Based on measurements of the degree day accumulation required for 50$\%$ of the egg population to hatch, we regard the duration of the non-diapause and post-diapause egg stages as 240.3 degree days relative to a base development threshold of $10.4^{\circ}$C [D. Calvin, personal correspondence].  (Here, the base development threshold refers to the minimum temperature required for any development to occur at all.)  Thus, for these stages (non-diapause and post-diapause eggs), $a$ may be obtained through the relation $a = \tilde{a}/240.3$, where $\tilde{a}$ represents developmental age in units of degree days.  The motile stage then begins at the moment of accumulation of 240.3 degree days, and is truncated (for modeling purposes) at a developmental age at which individuals have essentially ceased to reproduce due to exhaustion of metabolic resources.  From field data indicating that an accumulation of 1668.7 degrees days is required for 90$\%$ of individuals to reach oviposition [D. Calvin, personal correspondence], as well as the assumption that a female lays on average 3 clutches [T. Leskey, personal correspondence] spaced $\sim$100 degree days apart, we set the duration of the motile stage to $1668.7-240.3+200 = 1628.4$ degree days.  As lanternflies do not develop during diapause, advancement through this stage is not measured in degree days.  Instead, progression through diapause is measured in units referred to here as \textit{cold credits}; based on experimental evidence indicating that an egg kept at $0^o$C--$10^o$C will complete diapause in approximately $60$ days~\cite{SHIM2015}, we define one \textit{cold credit} to be the amount by which an individual advances through diapause in one day if kept in this temperature range.  Then, an accumulation of 60 cold credits is required for diapause termination, and the age variable $a$ may be defined for diapausing eggs through the relation $a = \bar{a}/60$, where $\bar{a}$ is the number of cold credits accumulated.

Development speeds in SLF are quantified by the degree day function, through which each temperature, $T$, is assigned a certain number of degree days by which an individual ages in one day of constant exposure to that temperature.  The advection coefficients, $\nu^i(T)$, have the common piecewise-linear form of the degree day function~\cite{SMYERS2021} for $i \in I/\{d\}$:
\begin{align} \label{eq:degreeday}
\nu^i(T) L^i =
\begin{cases}
0 & \text{ for }\mbox{   } T < T_{\nu^i,1},\\
\frac{\nu^{i,\textrm{max}}}{T_{\nu^i,2}-T_{\nu^i,1}}\left(T - T_{\nu^i,2}\right)+\nu^{i,\textrm{max}} & \text{ for } \mbox{   }T_{\nu^i,1} \leq T < T_{\nu^i,2},\\
\nu^{i,\textrm{max}} & \text{ for } \mbox{   }T \geq T_{\nu^i,2}.
\end{cases} 
\end{align}
Here, $T_{\nu^i,1} = 10.4^o$C is the base development threshold, and $T_{\nu^i,2} = 30^o$C is the minimum temperature at which the maximum developmental speed ($\nu^{i,\textrm{max}} = 19.6$ degree days/day) is reached [D. Calvin, personal correspondence]. The factor $L^i$ in Eq.~\eqref{eq:degreeday} reflects the rescaling to the unitless age variable, with $L^u = L^p = 240.3$ degree days and $L^b = 1628.4$ degree days.  For $i = d$, the advection coefficient has the form, 
\begin{align} \label{eq:diapauseadvection}
\nu^d(T) =
\begin{cases}
\nu^{d,\textrm{min}}  &\text{ for }\mbox{    }  T < T_{\nu^d,1}, \\
\left(\frac{\nu^{d,\textrm{max}}-\nu^{d,\textrm{min}}}{T_{\nu^d,2}-T_{\nu^d,1}} \right)(T-T_{\nu^d,1})+\nu^{d,\textrm{min}} & \text{ for }\mbox{   } T_{\nu^d,1} \leq T < T_{\nu^d,2},\\
\nu^{d,\textrm{max}} & \text{ for } \mbox{   }T_{\nu^d,2} \leq T < T_{\nu^d,3},\\
\left(\frac{\nu^{d,\textrm{min}}-\nu^{d,\textrm{max}}}{T_{\nu^d,4}-T_{\nu^d,3}} \right)(T-T_{\nu^d,3})+\nu^{d,\textrm{max}} & \text{ for } \mbox{   }T_{\nu^d,3} \leq T < T_{\nu^d,4},\\
\nu^{d,\textrm{min}} & \text{ for } \mbox{   }T \geq T_{\nu^d,4}.
\end{cases} 
\end{align}
In Eq.~\eqref{eq:diapauseadvection}, $(T_{\nu^d,2},T_{\nu^d,3})$ represents the interval over which advection through diapause is fastest.  By construction of the age variable, $a$, in the diapause domain (described above), we take $(T_{\nu^d,2},T_{\nu^d,3}) = (0^o\mbox{C},10^o\mbox{C})$ and $\nu^{d,\textrm{max}} = 1/60 \mbox{  }\text{day}^{-1}$.  Diapause advection is slowest in the intervals $(-\infty,T_{\nu^d,1}) = (-\infty,-5^o\mbox{C})$, $(T_{\nu^d,4},\infty) = (15^o\mbox{C},\infty)$, with $\nu^{d,\textrm{min}} = 1/300 \mbox{ } \text{day}^{-1}$.  The functions $m^i(T)$ for $i \in I/\{b\}$ have the form,
\begin{align} \label{eq:eggdeath}
m^i(T) =
\begin{cases}
s^{m^i}(T-T_{m^i,1}) & \text{ for }\mbox{   } T < T_{m^i,1},\\
0 & \text{ for } \mbox{   } T \geq T_{m^i,1},
\end{cases}
\end{align}
where $T_{m^i,1}$ is the threshold temperature below which temperature-induced mortality is induced in eggs.  Here, $s^{m^i} < 0$, so that mortality rates increase linearly with decreasing temperature.  The calibrated function parameters are based on data from~\cite{PARK2015} and~\cite{KEENA2021} and summarized in Table~\ref{table:parameters}.  We expect an $\sim 40 \%$ basal mortality rate for eggs, and therefore take $\alpha = 0.6$ in Eq.~\eqref{eq:fluxbalance1}~\cite{PARK2015,KEENA2021}.  The mortality function for motiles has the form $m^b(T) = -\ln(\max\{r(T),\tau\})/17$, where,
\begin{align} \label{eq:motiledeath}
r(T) =
\begin{cases}
s^{b,1}(T-T_{m^b,1}) + w & \text{ for }\mbox{   } T < T_{m^b,1},\\
w & \text{ for }\mbox{   } T_{m^b,1} \leq T < T_{m^b,2},\\
s^{b,2}(T-T_{m^b,2}) + w & \text{ for } \mbox{   } T \geq T_{m^b,2},
\end{cases} 
\end{align}
with $-\ln(w)/17$ the basal death rate to which motiles are subjected within the ideal temperature range $(T_{m^b,1},T_{m^b,2})$, and $0 < \tau \ll 1$.  We assume that $s^{b,1} > 0$ and $s^{b,2} < 0$, so that death rates increase as temperatures depart further from the ideal range.  The function $r(T)$ is a piecewise linear function fit to data from a 17-day survival study, with $r(T)$ representing the fraction of the starting population that survived a 17-day period kept at constant temperature $T$~\cite{KREITMAN2021}.  Assuming that at fixed temperatures death decreases the population exponentially, $m^b(T) = -\ln(\max\{r(T),\tau\})/17$ represents the exponential decay rate at temperature $T$.

Finally, the egg-laying kernel, $k(a)$, in Eq.~\eqref{eq:fluxbalance1} has the form,
\begin{align}
k(a) = 
\begin{cases}
0 &\mbox{ for } \mbox{  } a < a_{\textrm{r}},\\
\frac{\beta\exp\left(\frac{-(a-a_r)}{\gamma}\right)}{\gamma \left(1 - \exp\left(\frac{-(1-a_r)}{\gamma}\right)\right)} &\mbox{ for } \mbox{  } a\geq a_{\textrm{r}},
\end{cases}
\end{align}
where $\beta$ is the total number of female eggs laid by a female lanternfly over the course of her life, $\gamma$ is the average amount by which a female has aged between successive clutches, and $a_{\textrm{r}}$ is the average age of a female when she lays her first clutch.  (Both $\gamma$ and $a_{\textrm{r}}$ are given in the unitless age variable, $a$.)  We take $\beta = 50$ based on the idea that a female lays $\sim 3$ clutches [T. Leskey, personal correspondence] of $~30 - 50$ eggs each~\cite{SHIM2015}, about half of which are female.  Assuming a spacing of $\sim 100$ degree days between clutches, we take $\gamma = 100/L^b = 100/1628.4 \approx 0.061$.  Based on field data that indicates that 50$\%$ of females reach egg-laying after 1616.4 degree days, we take $a_{\textrm{r}} = (1616.4-L^u)/L^b = (1616.4-240.3)/1628.4 \approx 0.85$ [D. Calvin, personal correspondence].
\begin{table}[h!]
\begin{center}
 \begin{tabular}{|c|c|c|c|c|} 
 \hline
 Eqn. & Parameter & Meaning & Value & Ref.\\ 
 \hline\hline
  \ref{eq:degreeday} & $L^u, L^p$ & sessile development duration & 240.3 degree days & D. Calvin\\ 
 \hline
 \ref{eq:degreeday} & $L^b$ & motile development duration & 1628.4 degree days & D. Calvin\\ 
 \hline
 \ref{eq:degreeday} & $T_{\nu^i,1}, i \neq d$ & base development threshold  & 10.4$^o$C & D. Calvin\\ 
 \hline
 \ref{eq:degreeday} & $T_{\nu^i,2}, i \neq d$ & peak development threshold  & 30$^o$C & D. Calvin\\
 \hline
 \ref{eq:degreeday} & $\nu^{i,\textrm{max}}, i \neq d$ & max. development rate & 19.6 degree days & D. Calvin\\
 \hline
 \ref{eq:diapauseadvection} & $(T_{\nu^d,1},T_{\nu^d,4})$ & region of faster & (-5,15)$^o$C & \cite{SHIM2015}\\
  & & diapause advection & & \\
 \hline
 \ref{eq:diapauseadvection} & $(T_{\nu^d,2},T_{\nu^d,3})$ & region of max. & (0,10)$^o$C & \cite{SHIM2015}\\ 
 & & diapause advection & & \\
 \hline
 \ref{eq:diapauseadvection} & $\nu^{d,\textrm{max}}$ & max. diapause advection rate & 0.017 $\text{day}^{-1}$ & \cite{SHIM2015}\\
 \hline
 \ref{eq:diapauseadvection} & $\nu^{d,\textrm{min}}$ & min. diapause advection rate & 0.003 $\text{day}^{-1}$ & \cite{SHIM2015}\\
 \hline
 \ref{eq:eggdeath} & $T_{m^u,1}$ & cold-induced death threshold,  & 1.043$^o$C & \cite{PARK2015,KEENA2021,THOMAS2012}\\
 & & non-diapause eggs & &\\
 \hline
 \ref{eq:eggdeath} & $T_{m^i,1}, i = d,p$ & cold-induced death threshold,  & -3.957$^o$C & \cite{PARK2015,KEENA2021,THOMAS2012}\\
 & & diapause/post-diapause eggs & &\\
 \hline
 \ref{eq:eggdeath} & $s^{m^i}, i \neq b$ & cold-induced death rate, all eggs & -0.073 $\text{(day $\times\mbox{ }   ^o$C)}^{-1}$ & \cite{PARK2015,KEENA2021}\\
 \hline
 \ref{eq:motiledeath} & $(T_{m^b,1},T_{m^b,2})$ & ideal temp. range, motiles & (10,28.7)$^o$C & \cite{KREITMAN2021}\\
 \hline
 \ref{eq:motiledeath} & $-\ln(w)/17$ & basal death rate, motiles & 0.01 $\text{day}^{-1}$ & \cite{KREITMAN2021}\\
 \hline
 \ref{eq:motiledeath} & $s^{b,1}$ & cold-induced death rate, motiles & 0.156 $\text{(day $\times\mbox{ }^o$C)}^{-1}$ & \cite{KREITMAN2021}\\
 &  & diapause/post-diapause eggs & &\\
 \hline
 \ref{eq:motiledeath} & $s^{b,2}$ & heat-induced death rate, motiles & -0.072 $\text{(day $\times\mbox{ }^o$C)}^{-1}$ & \cite{KREITMAN2021}\\
 & & diapause/post-diapause eggs & &\\
 \hline
 \ref{eq:genpde} & $\sigma^{i}$, $i = u,p$ & age diffusion coefficient & $6.4 \times 10^{-4}$ & D. Calvin\\
 \hline
 \ref{eq:genpde} & $\sigma^i$, $i = b$ & age diffusion coefficient & $0.005$ & D. Calvin\\
 \hline
\end{tabular}
\caption{Summary of Calibrated Parameter Values}
\label{table:parameters}
\end{center}
\end{table}

\section{Numerical Method}
\label{sec:numericalmethod}

To compute the solution of Eqs.~\eqref{eq:genpde}--\eqref{eq:fluxbalance3} forward in time, we employ a finite volume framework to ensure conservation of mass in the absence of death and birth.  We formulate an operator splitting technique, updating the solution at each time step with respect to the advective and diffusive differential operators, and the reaction term, in Eq.~\eqref{eq:genpde} separately and sequentially.  Below, we devise a novel moving mesh method for the advection operation, with which the computed solution is robust under changes to the numerical parameters, even for solutions that are strongly peaked (that is, resemble a nascent delta function) due to diapause effects. This contrasts classical finite volume methods, in which solutions are afflicted by numerical diffusion or other deformations.  We first describe our moving mesh method on a single finite domain, and then describe the full operator splitting technique for solving Eqs.~\eqref{eq:genpde}--\eqref{eq:fluxbalance3} in detail.

\subsection{A moving mesh method}
\label{sec:movingmesh}

Here, we develop a moving mesh method to compute the solution, $\phi(a,t)$, of the one-dimensional advection equation,
\begin{align} \label{eq:simpleadvection}
\phi_t + \eta(t) \phi_a = 0 \mbox{  } \text{for } (a,t) \in (0,1) \times (0,t^*),
\end{align}
subject to the initial condition $\phi(a,0) = \phi_0(a)$.  The method is based on the simple idea that if $\mu = \int_t^{t+\Delta t} \eta(s) \textrm{d}s$ is the true advective shift in the solution $\phi$ over the time interval $[t,t+\Delta t]$, then $\int_{a_1}^{a_2} \phi(a,t) \textrm{d}a = \int_{a_1+\mu}^{a_2+\mu} \phi(a,t+\Delta t) \textrm{d} a$.  At each time step, $t^n$, the $a$-domain is discretized with respect to a mesh of cells with boundaries defined by a fixed partition $\{a_j\}$ of $[0,1]$ and a varying parameter, $\mu^n \geq 0$, by which the cells are shifted; $\mu^n$ represents the cumulative advective shift in the solution modulo the numerical cell width.  The numerical solution at $t^n$ then consists of a state vector, $\boldsymbol{\phi}^n$, containing the total mass ($\int \phi \mbox{  }\textrm{d}a$) in each numerical cell, coupled with the parameter $\mu^n$, which tracks the path of the cell masses relative to the unshifted reference mesh.

With this setup, the advancement in time is as follows: In Eq.~\eqref{eq:simpleadvection}, $t^* > 0$ denotes the chosen final time of the simulation; letting $\Delta t$ denote the time step size, we compute approximations to the solution $\phi$ at time steps $t^n = n\Delta t$, with $t^* = n^* \Delta t$ for some $n^* \in \mathbb{N}$.  To discretize the $a$-domain, we first denote by $a_j$ for $j = 0, 1, \ldots, N-1$ the partition points $a_j = j \Delta a$, with $\Delta a$ the chosen cell width, and $N \Delta a = 1$.  For any $\mu \geq 0$, we define a mesh, $C_{\mu}$, consisting of numerical cells, $C_{\mu,j}$ for $j = 0, 1, \ldots, N$, having the form,
\begin{align}
C_{\mu,j} =
\begin{cases}
[a_j,\mu \Delta a]  &\text{ for }\mbox{    }  j = 0, \\
[a_{j-1} + \mu \Delta a, a_j + \mu \Delta a] & \text{ for }\mbox{   } j = 1, \ldots, N-1,\\
[a_{j-1}+\mu \Delta a,1] & \text{ for } \mbox{   } j = N.
\end{cases} \label{eq:numericalcells1D}
\end{align}
The numerical solution at time step $t^n$ is defined by a pairing, $(\mu^n,\boldsymbol{\phi}^n)$, of a scalar $0 \leq \mu^n < 1$ and a state vector $\boldsymbol{\phi}^n = [\phi_0^n, \phi_1^n, \ldots, \phi_N^n]$, in which the cell-integrals
\begin{align}
\phi_j^n \approx \int_{C_{\mu^n,j}} \phi(a,t^n) \mbox{ } \textrm{d}a
\end{align}
form the numerical approximation. As with a traditional finite volume method, a given mesh shift and arrangement of cell-integrals can be associated with a reconstructed function $\tilde{\phi}(a,t^n)$ that is piecewise constant on the cells $C_{\mu^n,j}$ (Fig.~\ref{fig:piecewise_constant_approximation}) and satisfies
\begin{align}
\int_{C_{\mu^n,j}} \tilde{\phi}(a,t^n) \mbox{ } \textrm{d} a = \phi_j^n \mbox{ } \text{for} \mbox{ } j = 0, 1, \ldots, N.
\end{align}
The scalar $\mu^n$ measures temporally local advective shifts in the solution; setting $\mu^0 = 0$, it is updated iteratively by,
\begin{align} \label{eq:basicmu}
\mu^{n} =
\begin{cases}
\mu^{n-1} + \bar{\mu}^{n}  &\text{ if }\mbox{    }  \mu^{n-1} + \bar{\mu}^{n} < 1, \\
\mu^{n-1} + \bar{\mu}^{n} - 1 & \text{ if }\mbox{   } \mu^{n-1} + \bar{\mu}^{n}\geq 1,
\end{cases}
\end{align}
where $\bar{\mu}^{n} = \frac{\eta(t^{n-1}) \Delta t}{\Delta a}$.  Approximating the advective shift that occurs between time $t^{n-1}$ and $t^{n}$ as $\int_{t^{n-1}}^{t^{n}} \eta(t) \mbox{ } \textrm{d}t \approx \eta(t^{n-1}) \Delta t$, $\bar{\mu}^{n}$ is the fraction of one numerical cell's width by which the solution shifts between $t^{n-1}$ and $t^{n}$.  From Eq.~\eqref{eq:basicmu}, $\mu^n$ represents the cumulative advective shift in the solution since time $t^0$, modulo $\Delta a$.  At each $t^n$, the scheme used to compute $\boldsymbol{\phi}^n$ depends on the sign of $\mu^{n-1} + \bar{\mu}^n - 1$.  If $\mu^{n-1} + \bar{\mu}^n - 1 < 0$, then $\phi^n_j$ is computed as follows:
 \begin{align}
 \phi^{n}_j = 
 \begin{cases}
 \phi^{n-1}_j + \phi^{n}_{\textrm{in}} &\text{ for } \mbox{  } j = 0,\\
 \phi^{n-1}_j &\text{ for } \mbox{   } j = 1, \ldots, N-1,\\
 \phi^{n-1}_j - \phi^{n}_{\textrm{out}} &\text{ for } \mbox{   } j = N,
 \end{cases} \label{eq:basicschemeless1}
 \end{align}
where $\phi^{n}_{\textrm{out}} = \left(\frac{\bar{\mu}^{n}}{1-\mu^{n-1}}\right)\phi^{n-1}_{N}$ is the outflux through the boundary at $a = 1$ and $\phi^n_{\textrm{in}}$ is a prescribed influx through $a = 0$.  The piecewise constancy of the numerical solution, $\tilde{\phi}$, is used implicitly in the construction of $\phi^n_{\textrm{out}}$: the fraction of the mass ($\phi^{n-1}_N$) in the final numerical cell ($C_{\mu^{n-1},N}$) that exits through the boundary at $a=1$ simply equals the fraction of that numerical cell's width by which the solution shifts in that time step.  In this case in which $\mu^{n-1} + \bar{\mu}^n < 1$, the advective shift has not again surpassed $\Delta a$; per Eq.~\eqref{eq:basicschemeless1}, the scheme leaves the interior cell-integral values $\phi_j^n$ for $j = 1, \ldots, N-1$ unchanged, while adding mass to the boundary cell at $a=0$ and removing mass from the cell at $a=1$ in relation to the extent of the advective shift at the current time step.  Although the shift does not affect any explicit change in $\phi_j^n$ for $j = 1, \ldots, N-1$, it is accounted for in the underlying shifted mesh, $C_{\mu^n}$ (Fig.~\ref{fig:update_step}).
 
If instead $\mu^{n-1} + \bar{\mu}^n - 1 \geq 0$, then $\phi^n_j$ is computed as follows:
\begin{align}
 \phi^{n}_j = 
 \begin{cases}
 \left(\frac{\mu^{n}}{\mu^{n-1}}\right)\phi^{n}_{\textrm{in}} & \text{ for } \mbox{   } j = 0,\\
 \phi^{n-1}_{j-1} + \left(1 - \frac{\mu^{n}}{\mu^{n-1}}\right) \phi^{n}_{\textrm{in}}& \text{ for } \mbox{   } j = 1,\\
 \phi^{n-1}_{j-1} & \text{ for } \mbox{   } j = 2,\ldots, N-1,\\
 (1-\mu^{n}) \phi^{n-1}_{j-1} & \text{ for } \mbox{   } j = N,
 \end{cases} 
 \label{eq:basicschemegtr1}
 \end{align}
with outflux $\phi^{n}_{\textrm{out}} = \mu^n \phi^{n-1}_{N-1} + \phi^{n-1}_N$ through $a = 1$.  In this case in which $\mu^{n-1} + \bar{\mu}^n \geq 1$, the advective shift has now exceeded $\Delta a$, and this is accounted for by applying a right-shift to $\boldsymbol{\phi}^n$, as in Eq.~\eqref{eq:basicschemegtr1}, and moving the mesh back by a factor of $\Delta a$, as in Eq.~\eqref{eq:basicmu}.  The influx $\phi^n_{\textrm{in}}$ is distributed among the two leftmost cells $C_{\mu^n,0}, C_{\mu^n,1}$, while mass is removed from the two rightmost cells $C_{\mu^n,N-1}, C_{\mu^n,N}$ to an extent dictated by $\mu^n$ (Fig.~\ref{fig:update_step}).
\begin{figure}[htbp]
    \centering
    \begin{tikzpicture}[>=stealth,domain=0:6,samples=100,scale=0.9]
        \draw[->,thin,black] (0,-0.3) -- (0,4) node[above] {$\phi(a,\cdot)$};
        \draw[->,thin,black] (-0.3,0) -- (7,0) node[right] {$a$};
        \draw[ultra thin,black] (6,-0.3) -- (6,{cos(6 r)+2});
        \draw[color=black,ultra thin,fill=blue!20] plot (\x,{cos(\x r)+2}) -- (6,0) -- (0,0) -- (0,{cos(0 r)+2});
        \draw[color=blue,thick] plot (\x,{cos(\x r)+2});
        \path (3.3,3.8) node {true solution};
        \path (6,-0.55) node {1};
        \path (0,-0.55) node {0};
    \end{tikzpicture}
    \begin{tikzpicture}[>=stealth,scale=0.9]
        \draw[->,thin,black] (0,-0.3) -- (0,4) node[above] {$\tilde{\phi}(a,\cdot)$};
        \draw[->,thin,black] (-0.3,0) -- (7,0) node[right] {$a$};
        \draw[ultra thin,black] (1,-0.3) -- (1,2.84);
        \draw[ultra thin,black] (2,-0.3) -- (2,2.07);
        \draw[ultra thin,black] (3,-0.3) -- (3,1.23);
        \draw[ultra thin,black] (4,-0.3) -- (4,1.1);
        \draw[ultra thin,black] (5,-0.3) -- (5,1.8);
        \draw[ultra thin,black] (6,-0.3) -- (6,2.68);
        \draw[color=black,thin,fill=blue!20] (0,0) -- (1,0) -- (1,2.84) -- (0,2.84) -- cycle;
        \draw[thick,blue] (0,2.84) -- (1,2.84);
        \draw[color=black,thin,fill=blue!20] (1,0) -- (2,0) -- (2,2.07) -- (1,2.07) -- cycle;
        \draw[thick,blue] (1,2.07) -- (2,2.07);
        \draw[color=black,thin,fill=blue!20] (2,0) -- (3,0) -- (3,1.23) -- (2,1.23) -- cycle;
        \draw[thick,blue] (2,1.23) -- (3,1.23);
        \draw[color=black,thin,fill=blue!20] (3,0) -- (4,0) -- (4,1.1) -- (3,1.1) -- cycle;
        \draw[thick,blue] (3,1.1) -- (4,1.1);
        \draw[color=black,thin,fill=blue!20] (4,0) -- (5,0) -- (5,1.8) -- (4,1.8) -- cycle;
        \draw[thick,blue] (4,1.8) -- (5,1.8);
        \draw[color=black,thin,fill=blue!20] (5,0) -- (6,0) -- (6,2.68) -- (5,2.68) -- cycle;
        \draw[thick,blue] (5,2.68) -- (6,2.68);
        \path (6,-0.55) node {\small $1$};
        \path (0,-0.55) node {\small $a_0(=0)$};
        \path (1,-0.55) node {\small $a_1$};
        \path (2,-0.55) node {\small $a_2$};
        \path (3,-0.55) node {$\cdots$};
        \path (3.3,3.8) node[align=center] {numerical\\approximation};
        \path (4.5,-0.3) node {\small $\Delta a$};
        \draw[color=black,ultra thick] (4,-0.1) -- (4,0.1);
        \draw[color=black,ultra thick] (4,0) -- (5,0);
        \draw[color=black,ultra thick] (5,-0.1) -- (5,0.1);
    \end{tikzpicture}
    \caption{\textbf{Piecewise constant numerical approximation.} At any time $t$, the true function, $\phi(a,\cdot)$, is approximated by another function, $\tilde{\phi}(a,\cdot)$, piecewise constant on the cells of the moving mesh.  Here, $\mu = 0$ and the piecewise constant approximation is constructed with respect to the reference grid determined by the partition $\{a_j\}_{j=0}^{N-1}$ of the interval $[0,1]$.}
    \label{fig:piecewise_constant_approximation}
\end{figure}
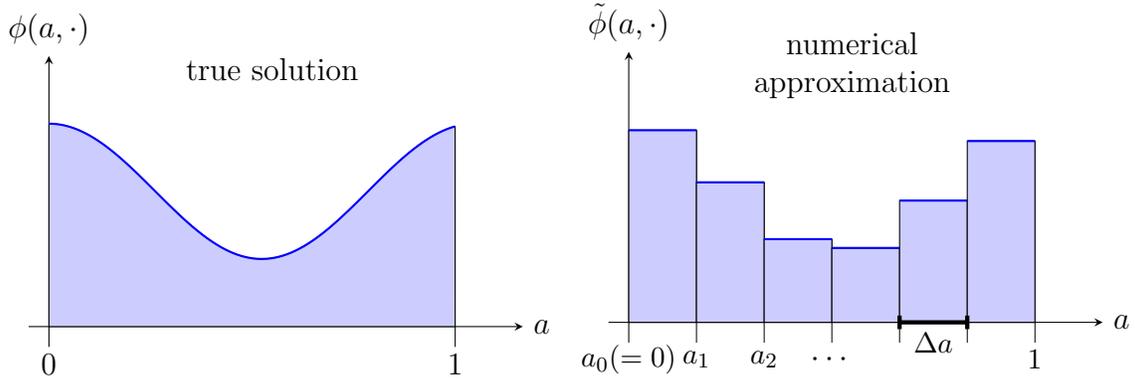
\begin{figure}[htbp]
\centering
        \begin{tikzpicture}[>=stealth,scale=0.7]
        \draw[->,thin,black] (0,-0.3) -- (0,4) node[above] {$\tilde{\phi}(a,t^{n-1})$};
        \draw[->,thin,black] (-0.3,0) -- (7,0) node[right] {$a$};
        \draw[ultra thin,black] (0.4,-0.3) -- (0.4,3.4);
        \draw[ultra thin,black] (1.4,-0.3) -- (1.4,2.84);
        \draw[ultra thin,black] (2.4,-0.3) -- (2.4,2.07);
        \draw[ultra thin,black] (3.4,-0.3) -- (3.4,1.23);
        \draw[ultra thin,black] (4.4,-0.3) -- (4.4,1.10);
        \draw[ultra thin,black] (5.4,-0.3) -- (5.4,1.80);
        \draw[ultra thin,black] (6,-0.3) -- (6,2.68);
        \draw[color=black,thin,fill=blue!20] (0,0) -- (0.4,0) -- (0.4,3.4) -- (0,3.4) -- cycle;
        \draw[thick,blue] (0,3.4) -- (0.4,3.4);
        \draw[color=black,thin,fill=blue!20] (0.4,0) -- (1.4,0) -- (1.4,2.84) -- (0.4,2.84) -- cycle;
        \draw[thick,blue] (0.4,2.84) -- (1.4,2.84);
        \draw[color=black,thin,fill=blue!20] (1.4,0) -- (2.4,0) -- (2.4,2.07) -- (1.4,2.07) -- cycle;
        \draw[thick,blue] (1.4,2.07) -- (2.4,2.07);
        \draw[color=black,thin,fill=blue!20] (2.4,0) -- (3.4,0) -- (3.4,1.23) -- (2.4,1.23) -- cycle;
        \draw[thick,blue] (2.4,1.23) -- (3.4,1.23);
        \draw[color=black,thin,fill=blue!20] (3.4,0) -- (4.4,0) -- (4.4,1.10) -- (3.4,1.10) -- cycle;
        \draw[thick,blue] (3.4,1.10) -- (4.4,1.10);
        \draw[color=black,thin,fill=blue!20] (4.4,0) -- (5.4,0) -- (5.4,1.8) -- (4.4,1.8) -- cycle;
        \draw[thick,blue] (4.4,1.8) -- (5.4,1.8);
        \draw[color=black,thin,fill=blue!20] (5.4,0) -- (6,0) -- (6,2.68) -- (5.4,2.68) -- cycle;
        \draw[thick,blue] (5.4,2.68) -- (6,2.68);
        \path (3.3,3.8) node {$\mu^{n-1} = 0.4$};
        \path (3.9,-0.45) node {$\Delta a$};
        \path (0.15,-0.9) node {\small $0.4\Delta a$};
        \path (5.65,-0.9) node {\small $0.6\Delta a$};
        \draw[color=black,ultra thick] (3.4,-0.1) -- (3.4,0.1);
        \draw[color=black,ultra thick] (3.4,0) -- (4.4,0);
        \draw[color=black,ultra thick] (4.4,-0.1) -- (4.4,0.1);
        \draw[color=black,ultra thick] (5.4,-0.1) -- (5.4,0.1);
        \draw[color=black,ultra thick] (5.4,0) -- (6,0);
        \draw[color=black,ultra thick] (6,-0.1) -- (6,0.1);
        \draw[color=black,ultra thick] (0,-0.1) -- (0,0.1);
        \draw[color=black,ultra thick] (0,0) -- (0.4,0);
        \draw[color=black,ultra thick] (0.4,-0.1) -- (0.4,0.1);
        \draw[->,color=black,thick] (0.2,-0.1) -- (0.2,-0.7);
        \draw[->,color=black,thick] (5.7,-0.1) -- (5.7,-0.7);
        \draw[->,color=black,thick] (2.5,-1,) .. controls (2.5, -3) and (-3,-3) .. (-3,-5) node[sloped,above,midway] {$\mu^{n-1} + \bar{\mu}^{n} = 0.7$};
        \draw[->,color=black,thick] (4.5,-1,) .. controls (4.5,-3) and (10,-3) .. (10,-5) node[sloped,above,midway] {$\mu^{n-1} + \bar{\mu}^{n} = 1.2$};
    \end{tikzpicture}\\
    \begin{tikzpicture}[>=stealth,scale=0.7375]
        \draw[->,thin,black] (0,-0.3) -- (0,4) node[above] {$\tilde{\phi}(a,t^{n})$};
        \draw[->,thin,black] (-0.3,0) -- (7,0) node[right] {$a$};
        \draw[ultra thin,black] (0.7,-0.3) -- (0.7,3.4);
        \draw[ultra thin,black] (1.7,-0.3) -- (1.7,2.86);
        \draw[ultra thin,black] (2.7,-0.3) -- (2.7,2.07);
        \draw[ultra thin,black] (3.7,-0.3) -- (3.7,1.23);
        \draw[ultra thin,black] (4.7,-0.3) -- (4.7,1.10);
        \draw[ultra thin,black] (5.7,-0.3) -- (5.7,1.80);
        \draw[ultra thin,black] (6,-0.3) -- (6,2.68);
        \draw[color=black,thin,fill=green!20] (0,0) -- (0.3,0) -- (0.3,2) -- (0,2) -- cycle;
        \draw[thick,green] (0,2) -- (0.3,2);
        \draw[color=black,thin,fill=blue!20] (0.3,0) -- (0.7,0) -- (0.7,3.4) -- (0.3,3.4) -- cycle;
        \draw[thick,blue] (0.3,3.4) -- (0.7,3.4);
        \draw[color=black,thin,fill=blue!20] (0.7,0) -- (1.7,0) -- (1.7,2.86) -- (0.7,2.86) -- cycle;
        \draw[thick,blue] (0.7,2.86) -- (1.7,2.86);
        \draw[color=black,thin,fill=blue!20] (1.7,0) -- (2.7,0) -- (2.7,2.07) -- (1.7,2.07) -- cycle;
        \draw[thick,blue] (1.7,2.07) -- (2.7,2.07);
        \draw[color=black,thin,fill=blue!20] (2.7,0) -- (3.7,0) -- (3.7,1.23) -- (2.7,1.23) -- cycle;
        \draw[thick,blue] (2.7,1.23) -- (3.7,1.23);
        \draw[color=black,thin,fill=blue!20] (3.7,0) -- (4.7,0) -- (4.7,1.1) -- (3.7,1.1) -- cycle;
        \draw[thick,blue] (3.7,1.1) -- (4.7,1.1);
        \draw[color=black,thin,fill=blue!20] (4.7,0) -- (5.7,0) -- (5.7,1.8) -- (4.7,1.8) -- cycle;
        \draw[thick,blue] (4.7,1.8) -- (5.7,1.8);
        \draw[color=black,thin,fill=blue!20] (5.7,0) -- (6,0) -- (6,2.68) -- (5.7,2.68) -- cycle;
        \draw[thick,blue] (5.7,2.68) -- (6,2.68);
        \draw[color=black,thin,fill=blue!10] (6,0) -- (6.3,0) -- (6.3,2.68) -- (6,2.68) -- cycle;
        \draw[thick,blue] (6,2.68) -- (6.3,2.68);
        \path (3.3,3.8) node {$\mu^{n} = 0.7$};
        \path (4.2,-0.45) node {$\Delta a$};
        \path (0.35,-0.9) node {\small $0.7\Delta a$};
        \path (5.85,-0.9) node {\small $0.3\Delta a$};
        \draw[color=black,ultra thick] (3.7,-0.1) -- (3.7,0.1);
        \draw[color=black,ultra thick] (3.7,0) -- (4.7,0);
        \draw[color=black,ultra thick] (4.7,-0.1) -- (4.7,0.1);
        \draw[color=black,ultra thick] (5.7,-0.1) -- (5.7,0.1);
        \draw[color=black,ultra thick] (5.7,0) -- (6,0);
        \draw[color=black,ultra thick] (6,-0.1) -- (6,0.1);
        \draw[color=black,ultra thick] (0,-0.1) -- (0,0.1);
        \draw[color=black,ultra thick] (0,0) -- (0.7,0);
        \draw[color=black,ultra thick] (0.7,-0.1) -- (0.7,0.1);
        \draw[->,color=black,thick] (0.35,-0.1) -- (0.35,-0.7);
        \draw[->,color=black,thick] (5.85,-0.1) -- (5.85,-0.7);
        \draw[->,thick] (6.2,1) to [out=90,in=180] (6.8,1.7) node[right,align=center] {\small mass\\ \small exits\\ \small through\\ \small $a=1$};
        \draw[->,thick] (-0.5,2.7) to [out=0,in=90] (0.2,1.7);
        \path (-1,2.5) node[align=center] {\small mass\\ \small enters\\ \small first\\ \small cell};
    \end{tikzpicture}
        \begin{tikzpicture}[>=stealth,scale=0.7375]
        \draw[->,thin,black] (0,-0.3) -- (0,4) node[above] {$\tilde{\phi}(a,t^{n})$};
        \draw[->,thin,black] (-0.3,0) -- (7,0) node[right] {$a$};
        \draw[ultra thin,black] (0.2,-0.3) -- (0.2,2);
        \draw[ultra thin,black] (1.2,-0.3) -- (1.2,3.4);
        \draw[ultra thin,black] (2.2,-0.3) -- (2.2,2.84);
        \draw[ultra thin,black] (3.2,-0.3) -- (3.2,2.07);
        \draw[ultra thin,black] (4.2,-0.3) -- (4.2,1.23);
        \draw[ultra thin,black] (5.2,-0.3) -- (5.2,1.1);
        \draw[ultra thin,black] (6,-0.3) -- (6,1.8);
        \draw[color=black,thin,fill=green!20] (0,0) -- (0.2,0) -- (0.2,2) -- (0,2) -- cycle;
        \draw[thick,green] (0,2) -- (0.2,2);
        \draw[color=black,thin,fill=green!20] (0.2,0) -- (0.8,0) -- (0.8,2) -- (0.2,2) -- cycle;
        \draw[thick,green] (0.2,2) -- (0.8,2);
        \draw[color=black,thin,fill=blue!20] (0.8,0) -- (1.2,0) -- (1.2,3.4) -- (0.8,3.4) -- cycle;
        \draw[thick,blue] (0.8,3.4) -- (1.2,3.4);
        \draw[color=black,thin,fill=blue!20] (1.2,0) -- (2.2,0) -- (2.2,2.84) -- (1.2,2.84) -- cycle;
        \draw[thick,blue] (1.2,2.84) -- (2.2,2.84);
        \draw[color=black,thin,fill=blue!20] (2.2,0) -- (3.2,0) -- (3.2,2.07) -- (2.2,2.07) -- cycle;
        \draw[thick,blue] (2.2,2.07) -- (3.2,2.07);
        \draw[color=black,thin,fill=blue!20] (3.2,0) -- (4.2,0) -- (4.2,1.23) -- (3.2,1.23) -- cycle;
        \draw[thick,blue] (3.2,1.23) -- (4.2,1.23);
        \draw[color=black,thin,fill=blue!20] (4.2,0) -- (5.2,0) -- (5.2,1.1) -- (4.2,1.1) -- cycle;
        \draw[thick,blue] (4.2,1.1) -- (5.2,1.1);
        \draw[color=black,thin,fill=blue!20] (5.2,0) -- (6,0) -- (6,1.8) -- (5.2,1.8) -- cycle;
        \draw[thick,blue] (5.2,1.8) -- (6,1.8);
        \draw[color=black,thin,fill=blue!10] (6,0) -- (6.2,0) -- (6.2,1.8) -- (6,1.8) -- cycle;
        \draw[thick,blue] (6,1.8) -- (6.2,1.8);
        \draw[color=black,thin,fill=blue!10] (6.2,0) -- (6.8,0) -- (6.8,2.68) -- (6.2,2.68) -- cycle;
        \draw[thick,blue] (6.2,2.68) -- (6.8,2.68);
        \path (3.3,3.8) node {$\mu^{n} = 0.2$};
        \path (4.7,-0.4) node {\small $\Delta a$};
        \path (0.1,-0.9) node {\small $0.2\Delta a$};
        \path (5.5,-0.9) node {\small $0.8\Delta a$};
        \draw[color=black,ultra thick] (4.2,-0.1) -- (4.2,0.1);
        \draw[color=black,ultra thick] (4.2,0) -- (5.2,0);
        \draw[color=black,ultra thick] (5.2,-0.1) -- (5.2,0.1);
        \draw[color=black,ultra thick] (5.2,-0.1) -- (5.2,0.1);
        \draw[color=black,ultra thick] (5.2,0) -- (6,0);
        \draw[color=black,ultra thick] (6,-0.1) -- (6,0.1);
        \draw[color=black,ultra thick] (0,-0.1) -- (0,0.1);
        \draw[color=black,ultra thick] (0,0) -- (0.2,0);
        \draw[color=black,ultra thick] (0.2,-0.1) -- (0.2,0.1);
        \draw[->,color=black,thick] (0.1,-0.1) -- (0.1,-0.7);
        \draw[->,color=black,thick] (5.5,-0.1) -- (5.5,-0.7);
        \draw[->,thick] (6.1,1) to [out=90,in=180] (7.1,1.7) node[right,align=center] {\small mass\\ \small exits\\ \small through\\ \small $a=1$};
        \draw[->,thick] (-0.5,2.7) to [out=0,in=90] (0.5,1.7);
        \path (-1,2.5) node[align=center] {\small mass\\ \small enters\\ \small first\\ \small two\\ \small cells};
        \draw[->,thick] (-0.5,2.7) to [out=0,in=90] (0.1,1.4);
    \end{tikzpicture}
    \caption{\textbf{Update step of the numerical method, example shift sizes.}  At time step $t^{n-1}$, the cells of the mesh have been shifted to the right by $\mu^{n-1} \Delta a= 0.4 \Delta a$; each interior cell has width $\Delta a$, while the first and last are truncated (top).  Advancing the numerical solution from time step $t^{n-1}$ to $t^{n}$ is conducted by shifting the mesh to $\mu^{n} \Delta a$. The relative advective shift from $t^{n-1}$ to $t^{n}$ is $\bar{\mu}^{n}$.  The state of the leftmost cell(s) is updated based on the given influx through $a=0$, and the amount of mass leaving via the rightmost cell(s) is assigned as the domain outflux through $a=1$. If $\mu^{n-1} + \bar{\mu}^{n} < 1$, then $\mu^{n} = \mu^{n-1} + \bar{\mu}^{n}$ and the influx is added to the leftmost cell, while the outflux is removed from the rightmost cell (lower left). If $\mu^{n-1} + \bar{\mu}^{n} \geq 1$, then $\mu^{n} = \mu^{n-1} + \bar{\mu}^{n} - 1$, so that the current advective shift is applied ($+ \bar{\mu}^{n} \Delta a$) before the grid is shifted back by $\Delta a$.  The influx is added to the two leftmost cells, while the outflux is removed from the two rightmost cells (lower right).}
    \label{fig:update_step}
\end{figure}
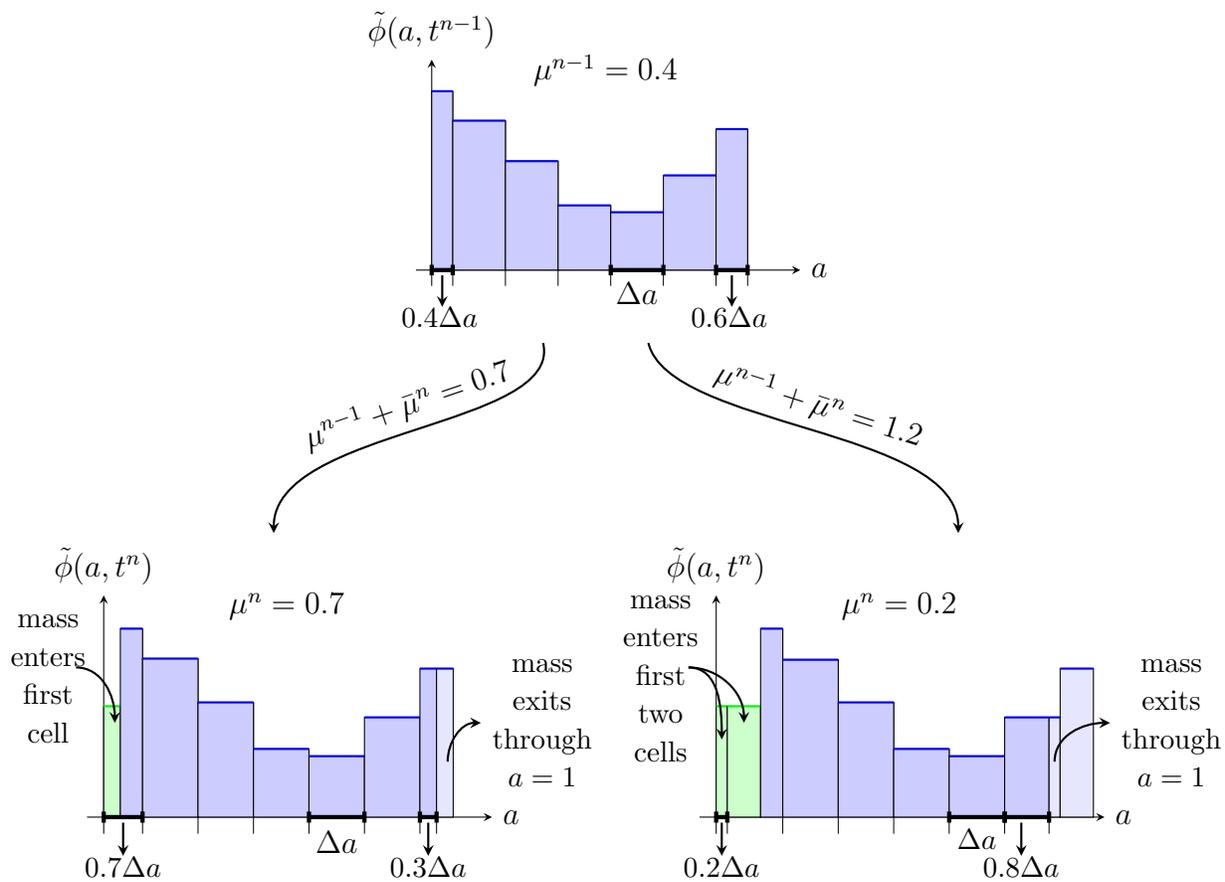

\subsection{Full numerical method}
\label{sec:fullmethod}
We now proceed to describe the full operator splitting method of solving Eqs.~\eqref{eq:genpde}--\eqref{eq:fluxbalance3}.  The method used to update the solution according to the advection operator is a direct extension of that detailed in Sec.~\ref{sec:movingmesh} in the case of one domain.  To begin, we denote by $\Delta t$ our time step size, and by $t^0 = 0$ our initial time, and compute the solution at times $t^n = t^0 + n \Delta t$ for $n \in \mathbb{N}$, until some prescribed final time, $t^*$.  As we are interested in computing eigenvalues of the one-year numerical solution operator, we typically choose $\Delta t$ so that $t^* = 365 = n^* \Delta t$ for some $n^* \in \mathbb{N}$.

To track advective shifts in the solutions as in Sec.~\ref{sec:movingmesh}, we discretize the age axis $a \in [0,1]$ in each of the four domains with respect to a moving mesh. To this end, we denote by $a^i_j$ for $j = 0, 1, \ldots, N^i-1$ and $i \in I$ the reference grid points $a^i_j = j\Delta a^i$, where $\Delta a^i$ is the chosen cell width in the $a$-domain corresponding to life stage $i$, and $N^i \Delta a^i = 1$.  For any $\mu^i \geq 0$, we define a mesh, $C_{\mu^i}$, consisting of numerical cells $C_{\mu^i,j}$ for $j = 0, 1, \ldots, N^i$ having the form,
\begin{align}
C_{\mu^i,j} =
\begin{cases}
[a^i_j,\mu^i \Delta a^i]  &\text{ for }\mbox{    }  j = 0, \\
[a^i_{j-1} + \mu^i \Delta a^i, a^i_j + \mu^i \Delta a^i] & \text{ for }\mbox{   } j = 1, \ldots, N^i-1,\\
[a^i_{j-1}+\mu^i \Delta a^i,1] & \text{ for } \mbox{   } j = N^i,
\end{cases} \label{eq:numericalcells}
\end{align}
which partition the age axis.  As in Sec.~\ref{sec:movingmesh}, the parameter $\mu^i$ is used to track the amount by which the solution in the domain corresponding to life stage $i$ has shifted due to advection, modulo $\Delta a^i$.

The numerical solution at time step $t^n$ is defined by a pairing, $(\boldsymbol{\mu}^n,\boldsymbol{\rho}^n)$, of the vectors $\boldsymbol{\mu}^n = [\mu^{u,n},\mu^{d,n},\mu^{p,n},\mu^{b,n}]$, where $0 \leq \mu^{i,n} < 1$ for all $i \in I$, and 
\begin{align}
\boldsymbol{\rho}^n = [\rho^{u,n}_0,\ldots,\rho^{u,n}_{N^u},\rho^{d,n}_0,\ldots,\rho^{d,n}_{N^d},\rho^{p,n}_0,\ldots,\rho^{p,n}_{N^p},\rho^{b,n}_0,\ldots,\rho^{b,n}_{N^b}]^T, 
\end{align}
with
\begin{align}
\rho^{i,n}_j \approx \int_{C_{\mu^{i,n},j}} \rho^i(a,t^n) \mbox{ }\textrm{d} a,
\end{align}
where $\rho^{i,n}_j$ is again based on a piecewise constant reconstruction $\tilde{\rho}^i(a,t^n)$.  (Fig.~\ref{fig:piecewise_constant_approximation}).  The vector $\boldsymbol{\mu}^n$ is determined by the temperature profile, $T(t)$, through the advection coefficients, $\nu^i$, in Eq.~\eqref{eq:genpde}, and is updated per a natural extension of Eq.~\eqref{eq:basicmu}.  The numerical solution is represented by total cell masses---as opposed to the cell averages that conventional finite volume methods use---in order to prevent numerical instabilities that occur when the boundary cells become very small, as is the case when $|\nu^i| \ll 1$.  To initialize the simulation, we take $\boldsymbol{\mu}^0 = [0,0,0,0]$ and define $\boldsymbol{\rho}^0$ with respect to the mesh $C_0$, setting,
\begin{align}
\rho^{i,0}_j = \int_{C_{0,j}} \rho_0^i(a) \mbox{ } \textrm{d}a,
\end{align}
for $i \in I$, $j = 0, \ldots, N^i$.  
The cell averages $\boldsymbol{\rho}^n$ are updated iteratively through the relation,
\begin{align} \label{eq:numericalupdate}
\boldsymbol{\rho}^{n} = \mathcal{R} \circ \mathcal{D} \circ \mathcal{A} \boldsymbol{\rho}^{n-1},
\end{align}
where $\mathcal{A}, \mathcal{D}, \mathcal{R}: \mathbb{R}^N \to \mathbb{R}^N$, with $N = 4 + \sum_{i=1}^4 N^i$, are the numerical advection, diffusion, and mortality operators, respectively.  In terms of its action on the components of the solution $\boldsymbol{\rho}^n$, the mortality operator satisfies,
\begin{align}
\mathcal{R}\rho^{i,n}_j  =  \exp{(-m^i(T(t^n))\Delta t)}\rho^{i,n}_j,
\end{align}
for $j = 0, \ldots, N^i$ and all $i \in I$.  (Due to the splitting method, neither $\mathcal{R}$ nor $\mathcal{D}$ affect the shift variables $\boldsymbol{\mu}^n$.) The diffusion operator implements a backward-time central-space finite difference scheme for the heat equation, with $\mathcal{D}\rho^{i,n}_j$ defined implicitly by,
\begin{align} \label{eq:diffusion}
-\theta \mathcal{D}\rho^{i,n}_{j-1} +  (1+2 \theta)\mathcal{D}\rho^{i,n}_j -\theta \mathcal{D}\rho^{i,n}_{j+1} = \rho^{i,n}_j
\end{align}
for $j = 1,\ldots,N^i-1$, and $i \in I/\{d\}$.  In Eq.~\eqref{eq:diffusion}, $\theta = \frac{\sigma^i \nu^i(T(t^n))\Delta t}{(\Delta a^i)^2}$. For simplicity, we assume no-diffusive-flux boundary conditions at $a \in \{0, 1\}$, taking $\mathcal{D}\rho^{i,n}_{j-1} - \mathcal{D}\rho^{i,n}_j = 0$ in Eq.~\eqref{eq:diffusion} for $j=0, N^i$ to append the system in Eq.~\eqref{eq:diffusion} with the equations,
\begin{align}
(1+ \theta)\mathcal{D}\rho^{i,n}_j -\theta \mathcal{D}\rho^{i,n}_{j+1} = \rho^{i,n}_j,
\end{align}
for $j = 0, N^i$ and $i \in I/\{d\}$.  The operator $\mathcal{A}$ is defined in analogue with the construction in Sec.~\ref{sec:movingmesh}.  Letting $\mu^{i,n} \geq 0$ represent the shift for life stage $i$ at time step $t^n$, we set $\mu^{i,0} = 0$ for all $i$, and define $\mu^{i,n}$ iteratively through the relation,
\begin{align} \label{eq:mu}
\mu^{i,n} =
\begin{cases}
\mu^{i,n-1} + \bar{\mu}^{i,n}  &\text{ if }\mbox{    }  \mu^{i,n-1} + \bar{\mu}^{i,n} < 1, \\
\mu^{i,n-1} + \bar{\mu}^{i,n} - 1 & \text{ if }\mbox{   } \mu^{i,n-1} + \bar{\mu}^{i,n}\geq 1,
\end{cases}
\end{align}
where $\bar{\mu}^{i,n} = \frac{\nu^i(T(t^{n-1})) \Delta t}{\Delta a^i}$.  The action of the advection operator at time $t^n$ is again determined by the sign of $\mu^{i,n-1} + \bar{\mu}^{i,n} - 1$.  If $\mu^{i,n-1} + \bar{\mu}^{i,n} - 1 < 0$, then $\mathcal{A}$ is represented in domain $i$ by the following scheme:
 \begin{align}
 \rho^{i,n}_j = \mathcal{A} \rho^{i,n-1}_j =
 \begin{cases}
 \rho^{i,n-1}_j + \rho^{i,n}_{\textrm{in}} &\text{ for } \mbox{  } j = 0,\\
 \rho^{i,n-1}_j &\text{ for } \mbox{   } j = 1, \ldots, N^i-1,\\
 \rho^{i,n-1}_j - \rho^{i,n}_{\textrm{out}} &\text{ for } \mbox{   } j = N^i,
 \end{cases} \label{eq:schemeless1}
 \end{align}
where $\rho^{i,n}_{\textrm{out}} = \left(\frac{\bar{\mu}^{i,n}}{1-\mu^{i,n-1}}\right)\rho^{i,n-1}_{N^i}$ and $\rho^{i,n}_{\textrm{in}}$ satisfies,
 \begin{align}
 \rho^{i,n}_{\textrm{in}} =
 \begin{cases}
 \xi(t^n) \rho^n_{\textrm{birth}} &\text{ for } \mbox{   } i = u,\\
  (1-\xi(t^n)) \rho^n_{\textrm{birth}} &\text{ for } \mbox{   } i = d,\\
 \rho^{d,n}_{\textrm{out}} &\text{ for } \mbox{   } i = p,\\
 \rho^{u,n}_{\textrm{out}} + \rho^{p,n}_{\textrm{out}} &\text{ for } \mbox{   } i = b,\\
 \end{cases} \label{eq:rhoin}
 \end{align}
where $\rho^n_{\textrm{birth}}$ is the total number of eggs laid between $t^{n-1}$ and $t^{n}$, and $\xi(t^n) = 1$ if $t^n$ falls after the winter solstice and before the summer solstice of the next year, and $\xi(t^n) = 0$ otherwise.  We approximate the total eggs laid from $t^{n-1}$ to $t^{n}$ as,
 \begin{align} \label{eq:rhobirth}
 \rho^n_{\textrm{birth}}= \sum_{j=1}^{N^b} \left[ K(a^b_{j} + \Delta a^b/2 + \mu^{b,n-1} + \bar{\mu}^{b,n}) - K(a^b_j + \Delta a^b/2 + \mu^{b,n-1})\right]\rho^{b,n-1}_j,
 \end{align}
with $K(a)$ defined as,
\begin{align}
K(a) =
\begin{cases}
\int_0^a k(x) \textrm{d} x &\mbox{ for } \mbox{  }a \leq 1,\\
\beta &\mbox{ for } \mbox{  }a > 1.
\end{cases}
\end{align}
Here, $k(a)$ is the birth kernel, per Eq.~\eqref{eq:fluxbalance1}, and we exploit the existing advective shift tracker, $\boldsymbol{\mu}^n$, to develop a semi-exact method for the birth term.  If, instead, $\mu^{i,n-1} + \bar{\mu}^{i,n} - 1 \geq 0$, $\mathcal{A}$ is then represented in domain $i$ with the following scheme,
 \begin{align}
 \rho^{i,n}_j = \mathcal{A}\rho^{i,n-1}_j =
 \begin{cases}
 \left(\frac{\mu^{i,n}}{\mu^{i,n-1}}\right)\rho^{i,n}_{\textrm{in}} & \text{ for } \mbox{   } j = 0,\\
 \rho^{i,n-1}_{j-1} + \left(1 - \frac{\mu^{i,n}}{\mu^{i,n-1}}\right) \rho^{i,n}_{\textrm{in}}& \text{ for } \mbox{   } j = 1,\\
 \rho^{i,n-1}_{j-1} & \text{ for } \mbox{   } j = 2,\ldots, N^i-1,\\
 (1-\mu^{i,n}) \rho^{i,n-1}_{j-1} & \text{ for } \mbox{   } j = N^i,
 \end{cases} \label{eq:schemegtr1}
 \end{align}
with $\rho^{i,n}_{\textrm{in}}$ defined as in Eq.~\eqref{eq:rhoin}.  Here, $\rho^{i,n}_{\textrm{out}} = \mu^{i,n}\rho^{i,n-1}_{N^i-1} + \rho^{i,n-1}_{N^i}$.  This formulation of the advection step yields accurate solutions, as the method is exact in the interiors of the domains and only approximate across the boundaries.  Moreover, the method is robust even if the advective speeds become very small or reach zero; if the advective speed in a given domain remains small for a period of time, the true solution will take the shape of a nascent delta function, while the numerical solution will accumulate mass in the leftmost cell of that domain.

Computations of $R_0$ in Sec.~\ref{sec:reproductivenumber} below will utilize mappings between numerical solutions at time steps spaced one year apart.  To frame this discussion, we first note that the mapping, $\bar{\mathcal{S}}_{n,\tilde{n}}$, that takes $(\boldsymbol{\mu}^n,\boldsymbol{\rho}^n) \to (\boldsymbol{\mu}^{\tilde{n}},\boldsymbol{\rho}^{\tilde{n}})$ is nonlinear due to the inclusion of the mesh shift parameter. If, however, $\boldsymbol{\mu}^n = \boldsymbol{\mu}^{\tilde{n}} = 0$, the mapping $\mathcal{S}_{n,\tilde{n}}$ that satisfies $\mathcal{S}_{n,\tilde{n}} \boldsymbol{\rho}^n = \boldsymbol{\rho}^{\tilde{n}}$ can be formulated with cell-integrals only, and is itself linear.  For the remainder of the text, we use the linear mapping $\mathcal{S}_{n,\tilde{n}}$ when computing $R_0$.  To work with this mapping, we can remap the solutions $\rho^{i,n}$ from ones defined with respect to the meshes $C_{\mu^{i,n}}$ to ones defined with respect to the mesh $C_0$.  To do this, we apply the following remapping to $\rho^{i,n}$:
 \begin{align} \label{eq:finalupdate}
 \rho^{i,n}_j \to 
 \begin{cases}
 0 &\text{ for } \mbox{   } j = 0,\\
 \rho^{i,n}_0 + (1-\mu^{i,n})\rho^{i,n}_1 &\text{ for } \mbox{   } j = 1,\\
 (1-\mu^{i,n})\rho^{i,n}_j + \mu^{i,n} \rho^{i,n}_{j-1} &\text{ for } \mbox{   } j = 2,\ldots,N^i-1,\\
 \mu^{i,n} \rho^{i,n}_{N^i-1} + \rho^{i,n}_{N^i} &\text{ for } \mbox{   } j = N^i,
 \end{cases}
 \end{align}
while setting $\mu^{i,n} = 0$. Note that this remapping is \textit{not} exact, and thus it is applied to the numerical solution only when strictly required, typically after the final step of the scheme to bring the system back to the starting mesh.

\section{Reproductive Number, $R_0$}
\label{sec:reproductivenumber}

To assess the capacity of the spotted lanternfly to establish in a given location, we begin by defining a notion of reproductive number (similarly, reproductive rate or net reproductive rate), $R_0$, appropriate to this species.  Reproductive numbers in viral spread modeling~\cite{NOWAK2000} and (net) reproductive rates in ecological population dynamics~\cite{STEINER2014} are typically defined with respect to generational changes as the number of individuals who become infected by one infected individual, and the number of offspring produced by one individual in her lifetime, respectively.  Although the spotted lanternfly typically has one generation per year (\textit{univoltinism}) in temperate climates~\cite{LEE2019}, multiple generations (\textit{multivoltinism}) are also possible in warmer regions.  Here, we are mostly concerned with annual temporal change independent of individual generations, thus we consider a year-to-year reproductive number associated with time $t$, defined as the number of individuals present at time $t (= t^0 + 365)$ per individual present at time $t^0$.  Throughout this work, we assume that each year has $365$ days.  Moreover, as we are presently interested in deriving a measure of the \textit{general} establishment potential intrinsic to a given location, we consider periodic temperature functions $T(t)$ that represent average conditions in that location.  For the study below, we consider a simplified temperature profile of the form,
\begin{align} \label{eq:temp}
T(t) = g \cos\left(\frac{2 \pi}{365}(t - \Phi)\right) + h,
\end{align}
where $h$, $g$, and $\Phi$ are, respectively, the mean, amplitude, and phase shift of the $365$-day periodic temperature function.  In the context of this one-year periodic temperature function, our reproductive number may naturally be associated with a day-of-year, as opposed to a particular date in history.  

To compute an approximation of this one-year reproductive number, we utilize the eigenvalue spectrum of the corresponding one-year numerical solution operator.  Based on the linear operator $\mathcal{S}_{n,\tilde{n}}$ described in Sec.~\ref{sec:fullmethod}, we first denote by $\mathcal{S}_n : \mathbb{R}^N \to \mathbb{R}^N$ the operator that satisfies $\mathcal{S}_n\boldsymbol{\rho}^n = \boldsymbol{\rho}^{\tilde{n}}$, where $(\tilde{n}-n)\Delta t = 365$.  From the periodicity of the temperature function in Eq.~\eqref{eq:temp}, it follows that $\mathcal{S}_{n} = \mathcal{S}_{n+365k}$ for $k \in \mathbb{N}$.  It therefore suffices to choose a start day of year and time of day, represented by $z$, and consider instead the one-year solution operator $\mathcal{S}_z : \mathbb{R}^N \to \mathbb{R}^N$ that satisfies, $\mathcal{S}_z \boldsymbol{\rho}^0 = \boldsymbol{\rho}^{n^*}$, using the temperature function $T(z+t)$ with $T$ as in Eq.~\eqref{eq:temp}.  (Here, $n^*$ satisfies $n^* \Delta t = 365$.)  As with the operator $\mathcal{S}_n$, $\boldsymbol{\rho}^0$ and $\boldsymbol{\rho}^{n^*}$ have been remapped to the mesh $C_0$.

We regard as an approximation to the $R_0$ associated with $z$ and $T(t)$ the dominant eigenvalue, $\lambda_1$, of the solution operator $\mathcal{S}_z$.  Owing to the linearity of the PDE and the moving mesh method, $\mathcal{S}_z$ is a linear operator; assuming it is also diagonalizable, 
$\lambda_1$ represents the inherent one-year growth factor asymptotically as $t \to \infty$.  The extent to which $\lambda_1$ alone captures transient population dynamical behavior depends on its relationship with subsequent eigenvalues: the smaller subsequent eigenvalues are in magnitude relative to $\lambda_1$, the more accurately $\lambda_1$ approximates $R_0$ for short times (depending on the initial condition).  As we are concerned with both short-term and long-term population establishment trends, we consider the first and second dominant eigenvalues in assessing $R_0$, as well as the transient behavior observed in simulations.

\subsection{Remarks on Numerical Method}

\begin{figure}
    \centering
    \includegraphics[width=\linewidth]{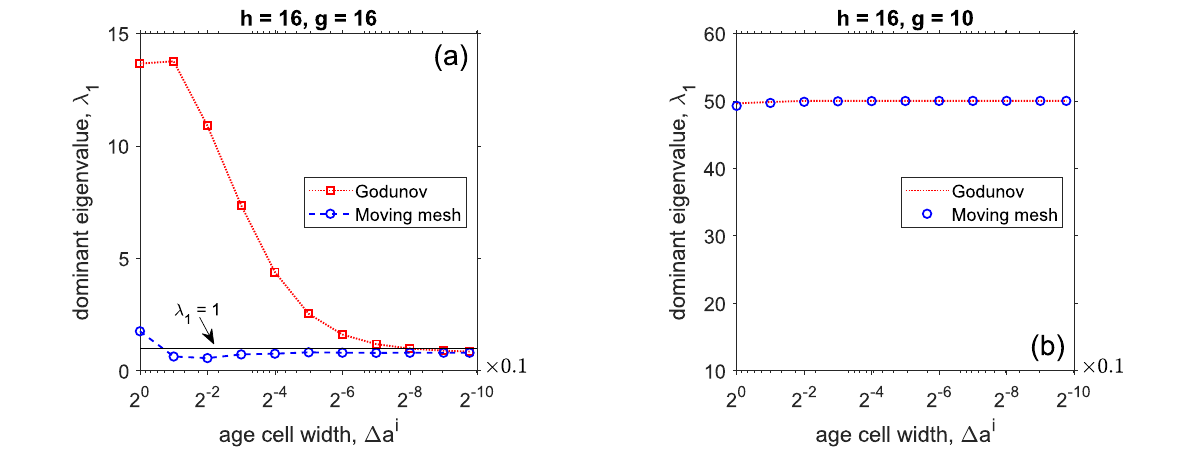}
    \caption{\textbf{Sensitivity of the dominant eigenvalue of the one-year solution operator to numerical parameters.} Computation of dominant eigenvalue, $\lambda_1$, as $\Delta a^i$ decreases.  In (a), $h = 16$, $g = 16$, while in (b), $h = 16$, $g = 10$.  In each case, each of the four domains is divided into cells of equal width, $\Delta a^i$, with $\Delta t$ chosen to satisfy the CFL condition given $\Delta a^i$.  The calibration described in Sec.~\ref{sec:calibration} is modified to eliminate basal egg and motile death, incorporate heat death in eggs, set $\sigma^i = 0$, and eliminate temperature mortality for temperatures within the range $6^o \textrm{C} - 26^o \textrm{C}$.  In (a), the Godunov method performs poorly until $\Delta a^i \lesssim (0.1) \times 2^{-8}$; in (b), the Godunov method produces accurate results at all discretizations.  In both (a) and (b), the moving mesh method produces accurate results at coarse discretizations.
    }
    \label{fig:eigenvaluesensitivity}
\end{figure}

The dominant eigenvalues and eigenvectors computed by the novel numerical method described in Sec.~\ref{sec:numericalmethod} are well-resolved
for moderate choices of the parameters $\Delta t$, $\Delta a^i$, and are robust with respect to small changes to those parameters. As the model under consideration tends to develop highly localized age-distributions, this is no longer the case if the advective numerical operator, $\mathcal{A}$ (Eq.~\eqref{eq:numericalupdate}), is replaced by a standard Godunov method. In that case, the eigenvalue computations may display a significant sensitivity to the choice of numerical parameters, as illustrated in Fig.~\ref{fig:eigenvaluesensitivity}.  In Fig.~\ref{fig:eigenvaluesensitivity}, we compare the dominant eigenvalue computed with the moving mesh and Godunov methods as the age cell width decreases for two different temperature profiles.  In these tests, we  take $\Delta a^i = \Delta a^j$ for all $i,j \in I$, for simplicity, and use a sinusoidal temperature profile (Eq.~\eqref{eq:temp}), fixing $\Phi = 203$ (July 22nd, 12 AM) with two mean/amplitude pairings, $h=16, g=16$ (a) and $h=16, g=10$ (b), as examples.  For testing purposes, we use an alternate parameterization of the model that (a) omits physiological diffusion ($\sigma^i = 0$ in Eq.~\eqref{eq:genpde}) to isolate the effects of numerical diffusion, (b) incorporates temperature-induced egg death (a feature that will be introduced into our model later, when data becomes available), (c) eliminates basal egg and motile death, and (d) boasts no temperature-induced death in the range $6^{\circ}\textrm{C}$--$26^{\circ}\textrm{C}$.  The latter two assumptions remove death as a population dynamical process entirely for temperature profile (b), creating conditions under which each individual reaches full egg-laying capacity as long as development rates are fast enough to move individuals through the egg-laying stage within one year.  As the advective shift parameter $\mu^{i,n}$ (Eqs.~\ref{eq:basicmu}, \ref{eq:mu}) is only present in the framework of the moving mesh method and not the Godunov method, we also replace the egg-laying term (Eq.~\ref{eq:rhobirth}) with an egg-laying flux using a semi-exact method for the integral in Eq.~\ref{eq:fluxbalance1}.

In Fig.~\ref{fig:eigenvaluesensitivity}(a), both eigenvalues approach the value $\lambda_1 \approx 0.8$ as $\Delta a^i$ decreases, however the moving mesh method (blue curve) does so much more rapidly than the Godunov method (red curve).  The Godunov method only achieves a result qualitatively consistent with the limiting value of $\approx 0.8$---that is, one for which $R_0 < 1$---for $\Delta a^i \lesssim (0.1) \times 2^{-8}$.  Simulations with such a small $\Delta a^i$ are inefficient, requiring significant computational time (hours in non-parallel implementations).  Moreover, if $\Delta a^i$ is chosen too large, this temperature profile may be mischaracterized as one for which $R_0 > 1$, resulting in a false assessment of establishment potential.  In turn, the eigenvalues computed with the moving mesh method (blue curve) are qualitatively accurate ($R_0 <1$) for discretizations with $\Delta a^i \lesssim (0.1) \times 2^{-1}$, and within a tolerance of $10^{-1}$ of the limiting value for $\Delta a^i = (0.1) \times 2^{-3}$.  With these discretizations, simulations require only seconds or minutes of computational time.  Under the temperature profile chosen used in (b), no individuals die and all are able to complete the egg-laying stage, thus we may intuit heuristically that the dominant eigenvalue should be $50$ (total per-capita lifetime egg-laying, $\beta$).  Clearly, in this example, both the moving mesh and Godunov methods compute the dominant eigenvalue accurately at even the coarsest discretizations.

When numerical sensitivity is observed in eigenvalue computations using the Godunov method, it is attributable to the numerical deformation effects inherent to fixed-grid finite volume methods (numerical diffusion in the Godunov case).  The age cell width directly affects the extent of the artificial diffusive spread in the age distribution, which in turn affects the timing and amount of egg-laying by motile adults.  When the solution is simulated using the moving mesh method, with the temperature profile from Fig.~\ref{fig:eigenvaluesensitivity}(a) and the dominant eigenvector as the initial condition, a strongly peaked solution forms, with the peak passing through the egg-laying stage just after temperatures reach their annual maximum ($32^{\circ}$C).  At these temperatures, newly-laid eggs perish due to heat exposure, yielding an $R_0 < 1$.  However, when the diffusive Godunov method is used instead, the strongly peaked solution spreads out, resembling a Gaussian.  In contrast with the moving mesh method, in which the cohort of motiles passes through the egg-laying stage in unison, individuals across the Gaussian enter the egg-laying stage at different times.  In particular, entry is delayed for many individuals to the left of the mean, allowing their egg-laying period to occur after egg-death-inducing heat has passed, leading to a higher $R_0 > 1$.  As $\Delta a^i$ decreases, the Gaussian distribution narrows, lessening the distortion of egg-laying rates and leading to eventual agreement with the results of the moving mesh method ($R_0 < 1$).  (In Fig.~\ref{fig:eigenvaluesensitivity}(b), the Godunov method captures egg-laying rates accurately at all discretizations because there is no source of death and all individuals reach full egg-laying capacity.)  Numerical diffusive effects therefore have significant potential to produce egg-laying rates that are fundamentally inaccurate and highly sensitive to numerical parameters.   The novel method described in Sec.~\ref{sec:numericalmethod} avoids the harmful effects of numerical diffusion by directly capturing the correct advection along characteristics, while maintaining the desirable finite volume fixed-width grid structure.

\subsection{Rank 1 Structure of $\mathcal{S}_z$ in Diapause}

In addition to its protective effects, a major benefit of diapause is its ability to synchronize the developmental ages of eggs in many distinct egg masses that were laid at different times.  This synchrony is advantageous later on, when individuals from different cohorts reach reproductive age roughly in unison, maximizing the potential to find suitable mates.  As diapause advection is dictated by ambient temperatures, the annual temperature profile determines how thoroughly diapause is able to synchronize the individuals in a population.  The climate most conducive to a high degree of synchrony is one in which the temperature function, $T(t)$, is characterized by a moderate mean and amplitude; this pairing results in a winter with temperatures sufficiently low to arrest development for long enough that all fall eggs may complete diapause before development resumes.  Under this assumption, all eggs laid in the fall complete diapause before any begin to develop, positioning them to begin development in unison once temperatures surpass the development threshold.  This action of diapause may be interpreted, mechanistically, as a \textit{correction} of the population's seasonal timing (\textit{phenology}): diapause adjusts the timelines of eggs laid at ecologically suboptimal times, so that they nevertheless rejoin their cohort and begin development at a more phenologically appropriate moment the following year, maximizing temperature-dependent survival and availability of food resources.

Temperature profiles that allow for diapause to exert this corrective influence typically involve cold winters with high rates of freezing-induced mortality in the system's more vulnerable agents.  The effects of phenological adjustment and cold mortality couple to result in the appearance of a low-rank structure of $\mathcal{S}_z$--that is, one in which $\mathcal{S}_z \approx \boldsymbol{\ell} \cdot \boldsymbol{r}$, for some vectors $\boldsymbol{\ell}, \boldsymbol{r}$, in which secondary eigenvalues are insignificant relative to the largest magnitude one.  The level of adherence to a low-rank structure is strong when $T(t)$ and the development and mortality functions satisfy the following criteria:
\begin{itemize}
\item (C1) The base development thresholds, $T_{\nu^i,1}$, for $i \in I/\{d\}$, are all equal, with their common value denoted by $T_{\nu,1}$,
\item (C2) The constants $g$ and $h$ in Eq.~\eqref{eq:temp} are chosen so that $\min_t T(t) < \min_{i \in I/\{d\}} T_{m^i,1}$.
\end{itemize}
Criterion (C2) guarantees that all sessile individuals will be subjected to cold-induced death for some portion of the year.  First consider the operator $\mathcal{S}_z$, where $z$ satisfies $T(z) = T_{\nu,1}$, $T'(z) < 0$, so that $z$ represents the moment at which development ceases (typically early winter).  Then denote by $t^m$ the final time step before development resumes, so that $t^m = \max\{t^n: t^n < \tilde{t}\}$, where $\tilde{t}$ is the point at which development resumes, satisfying $T(\tilde{t}) = T_{\nu,1}$, $T'(\tilde{t}) > 0$.  Then write $\mathcal{S}_z$ as $\mathcal{S}_z^1 \circ \mathcal{S}_z^0$, where $\mathcal{S}_z^0, \mathcal{S}_z^1 : \mathbb{R}^N \to \mathbb{R}^N$ are the operators that satisfy, respectively, $\mathcal{S}_z^0 \boldsymbol{\rho}^0 = \boldsymbol{\rho}^m$, $\mathcal{S}_z^1 \boldsymbol{\rho}^m = \boldsymbol{\rho}^{n^*}$.

\begin{proposition} \label{proposition:rank1}
If the diapause advection function, $\nu^d(T)$, is chosen to satisfy,
\begin{align} \label{eq:diapauseassumption}
    \int_0^{\tilde{t}} \nu^d(T(t))\;\mathrm{d}t > 1,
\end{align}
then there exists a subspace, $W$, of $\mathbb{R}^N$, such that:
\begin{enumerate}
    \item $\mathrm{dim}(\mathrm{im}(W)) = 1$,
    \item $\|\mathcal{S}_z^0 w^{\perp}\| \leq \|w^{\perp}\|$ for any $w^{\perp} \in W^{\perp}$,
\end{enumerate}
where $\mathrm{im}(W)$ is the image of $W$ under $\mathcal{S}_z^0$, $W^{\perp}$ is the orthogonal complement of W, and $\| \cdot \|$ is any $L_p$ norm.  Moreover, if Criteria (C1) and (C2) are satisfied, it follows that for any $\varepsilon > 0$, mortality functions $m^u$, $m^p$, and $m^b$ may be chosen such that,
\begin{align} \label{eq:lessthanvarepsilon}
    \|\mathcal{S}_z^0 w^{\perp}\| \leq \varepsilon \|w^{\perp}\|,
\end{align}
for any $w^{\perp} \in W^{\perp}$.
\end{proposition}

\textit{\textbf{Proof.}}  We begin by noting that, since $T(t^n) < T_{\nu,1}$ for all $t^n$ with $0 \leq n \leq m$, $\nu^i(T(t^n)) = 0$ for all such $t^n$.  In turn, $\theta = 0$, where $\theta$ is as defined in Eq.~\eqref{eq:diffusion}, implying that the operator $\mathcal{D}$ in Eq.~\eqref{eq:numericalupdate} is simply the identity operator while development is arrested, and may be ignored in considering $\mathcal{S}_z^0$.  

Now, denote by $e_k$ the standard basis vector of $\mathbb{R}^N$ with $1$ in the $k$-th component and $0$ elsewhere.  Then consider the subspace, $W$, defined as:
\begin{align}
    W := \{\textrm{span}\{e_k\}: e_k \cdot \boldsymbol{\rho}^n = \rho^{d,n}_j \text{ for some } j = 0, \cdots, N^d, \text{ or } e_k \cdot \boldsymbol{\rho}^n = \rho^{p,n}_0 \},
\end{align}
which spans the diapause portion of the numerical solution, as well as the first cell of the post-diapause domain.  Denote by $W^{\perp}$ the orthogonal complement of $W$.  Then, if Eq.~\eqref{eq:diapauseassumption} holds, there exist $\Delta t, \Delta a^d > 0$ sufficiently small that,
\begin{align} \label{eq:riemannsum}
    \Delta t \sum_{n=0}^m \nu^d(T(t^n)) > 1 + \Delta a^d,
\end{align}
if $t^n - t^{n-1} = \Delta t$ for all $n = 1,\ldots,m$, as Eq.~\eqref{eq:riemannsum} represents a Riemann sum approximation to the integral in Eq.~\eqref{eq:diapauseassumption}.  It follows that,
\begin{align} \label{eq:cflaccumulationlowerbound}
     \frac{\Delta t}{\Delta a^d} \sum_{n=0}^m \nu^d(T(t^n)) > \frac{1}{\Delta a^d} + 1 = N^d + 1.
\end{align}
As the expression on the left represents the total advective shift from $t^0$ through $t^m$ relative to $\Delta a^d$, Eq.~\eqref{eq:cflaccumulationlowerbound} indicates that there have been at least $N^d + 1$ right advective shifts in the diapause solution.  Moreover, $\nu^b(T(t^n)) = 0$ for $n = 0,\ldots,m$, thus $\rho^n_{\textrm{birth}} = 0$ from Eqs.~\eqref{eq:rhobirth} and~\eqref{eq:mu}, and therefore $\rho^{d,n}_0 = 0$ for $n = 1,\ldots,m$.  From this and the fact that at least $N^d + 1$ right shifts have occurred in the diapause domain, it follows that $\rho^{d,m}_j = 0$ for $j = 0,\ldots,N^d$.  Moreover, $\nu^p(T(t^n)) = 0$ for $n = 0,\ldots,m$, and thus $\rho^{p,m}_0 = \rho^{p,0}_0 + \sum_{j=0}^{N^d} \rho^{d,0}_j$.  In particular, $\textrm{im}(W)$ is spanned by the vector $e_j$ satisfying $e_j \cdot \boldsymbol{\rho}^m = \rho^{p,m}_0$, and $\textrm{dim}(\textrm{im}(W)) = 1$.  This establishes 1.

As the operator $\mathcal{A}$ in Eq.~\eqref{eq:numericalupdate} acts as the identity operator in all domains except the diapause domain, we have,
\begin{align}
    \rho^{i,m}_j = \left(\exp\left(-\Delta t \sum_{n=0}^m m^i(T(t^n))\right) \right)\rho^{i,0}_j,
\end{align}
for $i \in I/\{d\}$ and $j = 0,\ldots,N^i$ (except $\rho^{p,m}_0$), from which 2. immediately follows.  Eq.~\eqref{eq:lessthanvarepsilon} holds by then choosing $m^i$ for $i \in I/\{d\}$ sufficiently large to guarantee,
\begin{align}
   \max_{i \in I/\{d\}} \left(\exp\left(-\Delta t \sum_{n=0}^m m^i(T(t^n))\right) \right) < \varepsilon,
\end{align}
which is possible under Criterion (C2). \qed

In practice, the results of Prop.~\eqref{proposition:rank1} endow the operator $\mathcal{S}_z^0$ with a low-rank structure that naturally carries over to the full year operator $\mathcal{S}_z$.  The extent to which this low-rank structure is apparent depends heavily on the choice of $T(t)$:~the winter non-development period must present temperature conditions that allow all eggs to complete diapause before development resumes in order to achieve synchrony in the diapause domain.  The choice of $m^i(T)$ is also significant, as higher mortality rates that kill off other agents lead to smaller, less influential secondary modes in the eigenvalue spectrum of $\mathcal{S}_z$.  Here, the dominant eigenvector reflects the phenologically ideal age distribution at day of year/time of day $z$, while $\lambda_1$ represents the annual growth factor.  As the temperature profile departs from these assumptions, the degree of adherence to the low-rank structure also diminishes, and several modes may contribute to the population dynamical behavior.

\subsection{Numerical Results}

\begin{figure}
    \centering
    \includegraphics[width=\linewidth]{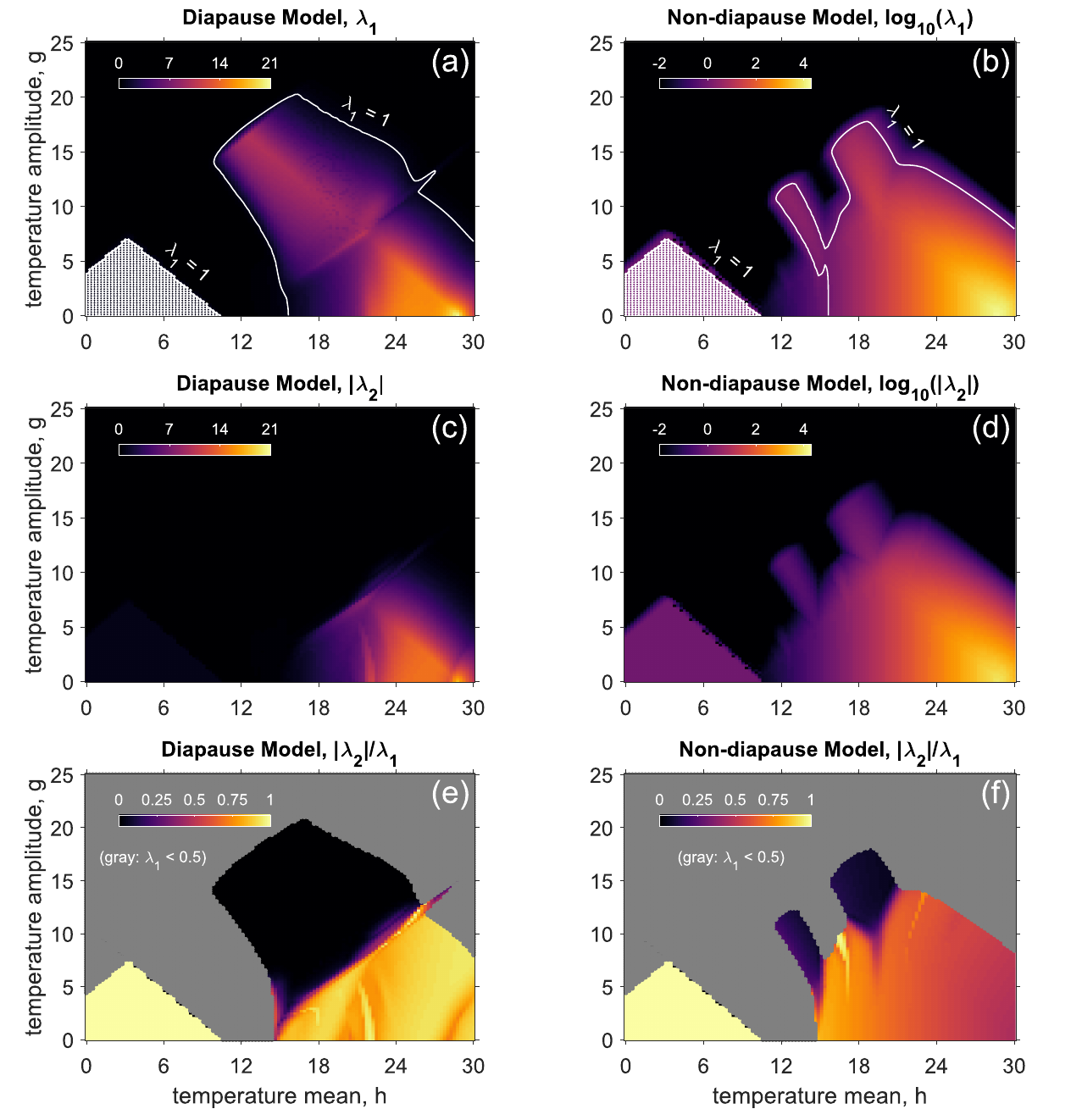}
    \caption{\textbf{Eigenvalues of the one-year numerical solution operator.}
    Plots of $\lambda_1$, $|\lambda_2|$, and $|\lambda_2|/\lambda_1$ under the diapause (a,c,e) and non-diapause (b,d,f) models.  (Note that (b,d) are plotted in log scale, the rest in linear scale.)  In the dark regions (a-d), climates are inhospitable to establishment.  In the high-mean/low-amplitude region (lower right, a-d), climates support population growth.  A low-rank region appears in the center of the parameter domain in diapause (a,c,e).  Low means and amplitudes (lower left, a-d) represent static laboratory conditions.  In the diapause case, $|\lambda_2|/\lambda_1 \approx 1$ in the high-mean/low-amplitude region (e), while the gap between $\lambda_1$ and $|\lambda_2|$ is larger in (f).  This figure was created in MATLAB using the inferno colormap (https://www.mathworks.com/matlabcentral/fileexchange/51986-perceptually-uniform-colormaps)}
    \label{fig:parametersweepdiapause}
\end{figure}

To explore the dependence of our estimated $R_0$ on the temperature profile in a given location, we conduct parameter sweeps in the parameters $g$ and $h$, representing amplitude and mean, in Eq.~\eqref{eq:temp}.  (In all computations presented below, we set the phase shift, $\Phi$, to $\Phi = 203$ (July 22nd, 12 AM), as it varies little from this value in our geographic region of interest.)  In Fig.~\ref{fig:parametersweepdiapause}, the first and second dominant eigenvalues, $\lambda_1$ and $\lambda_2$, of $\mathcal{S}_z$ are plotted, under the diapause (a,c) and non-diapause models (b,d), with $z = 203$ (July 22nd, 12 AM).  (This choice of $z$ is mostly arbitrary, as there is essentially no change in the spectrum as $z$ varies with a fixed temperature profile; thus, the spectrum is most naturally associated with the pairing $(g,h)$ and not a day of year.)  The ratio, $|\lambda_2|/\lambda_1$ is also plotted (e,f).  The functional forms of $\nu^i(T), m^i(T), k(a)$ are as described in Sec.~\ref{sec:calibration}.  (By virtue of its biological meaning, $\lambda_1$ is typically real and nonnegative under normal physiological assumptions and reasonable temperature profiles, but subsequent eigenvalues may be negative or complex.)  

Under the diapause model, the mean-amplitude phase space is partitioned into four qualitatively distinct regions (Fig.~\ref{fig:parametersweepdiapause}(a,c,e)).  The darkest areas in Fig.~\ref{fig:parametersweepdiapause}(a) correspond to extreme means and/or amplitudes, representing temperature profiles fundamentally unconducive to population establishment.  A low mean results in slow development rates, which may prevent motiles from reaching the egg-laying stage, while high amplitudes suggest periods of extreme heat and cold, leading to high mortality.  (The low-mean, low-amplitude area in the lower left corner represents a climate in which annual temperatures stay within the narrow range in which no physiological activity occurs:~temperatures are too low for development, but not low enough for cold-induced death.  In this region, the initial population neither ages nor dies, and does not lay eggs, thus the dominant eigenvalue is 1.  While this temperature profile is unrealistic in natural settings, it may represent certain laboratory conditions.  The unrealistic nature of this experimental stasis would contribute an additional source of mortality, which is, however, currently unstudied and thus beyond the scope of our paper.)  The purple/deep pink central region in Fig.~\ref{fig:parametersweepdiapause}(a) is identifiable as the region characterised by a low-rank structure, as $\lambda_1$ is as large as 10 in this region, while $|\lambda_2| \approx 0$ in the same region in Fig.~\ref{fig:parametersweepdiapause}(c).  Parameter combinations at the center of this region (deep pink) represent the strongest level of adherence to a low-rank structure.  Populations subjected to temperature profiles in the high-mean/low-amplitude region in the lower right corner of Fig.~\ref{fig:parametersweepdiapause}(a) have the strongest establishment potential, and may even experience multivoltinism.  In this region, temperatures are warm, leading to fast development rates and little cold-induced death.  Although individuals may go through diapause, the population's time evolution is not predominantly dictated by the diapause process; fast egg development rates compensate for time spent (perhaps unnecessarily) in diapause to yield eventual non-diapause egg-laying that triggers multivoltinism and explosive population growth.  The synchronizing effect of diapause observed in cooler temperature regions is absent, and complex second eigenvalues close in magnitude to the first (Fig.~\ref{fig:parametersweepdiapause}(c,e)) may produce temporally oscillatory fluctuations. 

Under the non-diapause model, the high-mean/low-amplitude region of the parameter domain again presents high $R_0$ values, while the low-rank area of the diapause domain mostly disappears.  In this case, the eigenvalues are significantly larger in the high-mean/low-amplitude region than in the diapause model, indicating higher annual growth rates in non-diapausing populations under reasonably warm temperature conditions (Fig~\ref{fig:parametersweepdiapause}(b)).  While this might naturally call the usefulness of diapause into question, one may note that $|\lambda_1| \ll 1$ in much of the central region of Fig~\ref{fig:parametersweepdiapause}(b) that belongs to the low-rank area with $|\lambda_1| > 1$ in Fig.~\ref{fig:parametersweepdiapause}(a).  In such climates, a population of eggs that passed through diapause would likely thrive, while a non-diapausing population would not.  Per Fig.~\ref{fig:parametersweepdiapause}(f), the gap in the magnitudes of $\lambda_1$ and $|\lambda_2|$ is much wider than in diapause (Fig.~\ref{fig:parametersweepdiapause}(e)).

\subsection{Discussion of Establishment Potential}
\label{sec:discussion}
In the context of our model, pest establishment potential in a given location is fully understood by considering both $R_0$, the asymptotic growth factor, and the transient behavior of an initial population.  For a population to establish, it must be the case that $R_0 > 1$ and that the initial population and some number of subsequent cohorts are able to each lay sufficient eggs while transitioning to the new location's phenology.  To explore transient behavior in locations with asymptotic establishment potential ($R_0 > 1$), we simulated Eqs.~\eqref{eq:genpde}-\eqref{eq:fluxbalance3} for a selection of temperature profiles from qualitatively distinct regions of the mean-amplitude parameter domain (Fig.~\ref{fig:parametersweepdiapause}). For each temperature profile, we computed the solution of the PDE system for a variety of initial conditions over a period of four years.  We used initial conditions representing the arrival of a cohort of 100 individuals of the same developmental age, $a_{\textrm{init}}$, on a certain day/time of year, $z_{\textrm{init}}$ ($\rho_0^i(a) = 100\,\delta(a-a_{\textrm{init}})$ for some $i \in I$, $\rho_0^i(a) \equiv 0$ otherwise).  This initial condition represents such scenarios as a clutch of eggs or a cohort of motiles arriving together in a new location; it has the advantage of simplicity, and allows us to test sensitivity to the developmental age of arriving individuals and seasonal timing.  The choice of an initial population size of 100 is, of course, arbitrary, as the PDE is linear.

Much of the sensitivity of the transient behavior can be classified in terms of the dominant eigenvalues, with great sensitivity observed when $\lambda_1 > 1$ and $|\lambda_2| \ll \lambda_1$ and little sensitivity when $\lambda_1 > 1$ and $|\lambda_2| \sim \lambda_1$.  Both of these configurations occur under the diapause model: the former results from temperature profiles in the low-rank region of the mean-amplitude domain, while the latter occurs in the high-mean/low-amplitude region (Fig.~\ref{fig:parametersweepdiapause}(a)). Under the non-diapause model, the high-mean/low-amplitude region of the parameter domain presents a median between these extremes:~overall magnitudes of $\lambda_1, \lambda_2$ are significantly larger than they are in diapause, but $|\lambda_2|$ is less comparable to $\lambda_1$.  Transient dynamics are potentially still sensitive to age and timing of arriving populations when $|\lambda_2| < 1$, as is the case in the ``boundary" area of this high-mean/low-amplitude region.  

\begin{figure}
    \centering
    \includegraphics[width=\linewidth]{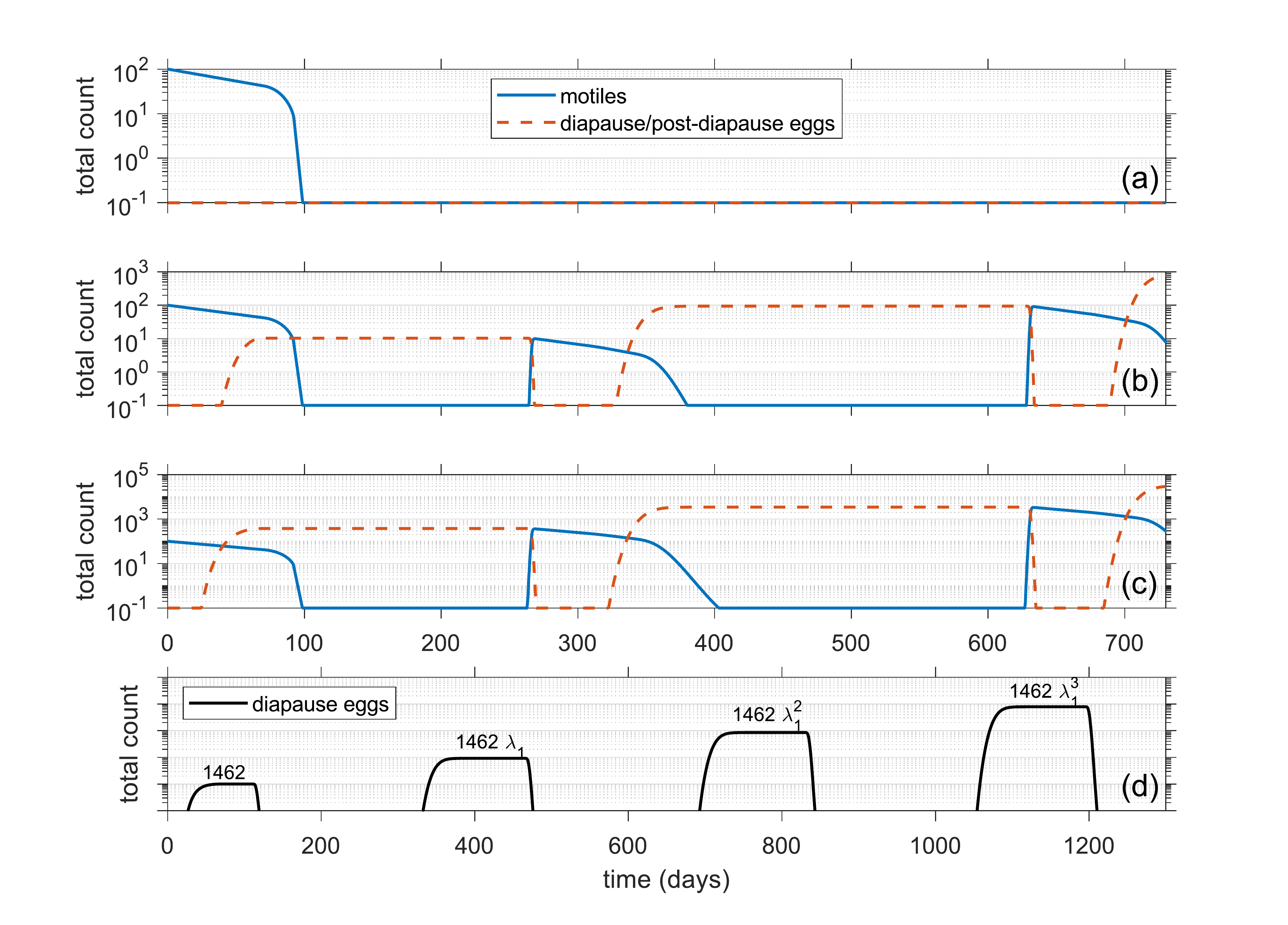}
    \caption{\textbf{Transient Behavior in the Diapause Rank 1 Region.}  $T(t)$ (Eq.~\eqref{eq:temp}) is parameterized by $h = 14.1$, $g = 15.5$, resulting in $\lambda_1 \approx 9.1$, $|\lambda_2| \approx 10^{-9}$.  In each case, the simulation begins on day of year 240 (August 28th, 12 AM).  In (a), 100 motiles of age $a_{\textrm{init}} = 0.11$ initialize the population, which quickly dies off.  In (b), 100 motiles of age $a_{\textrm{init}} = 0.36$ initialize the population, going on to lay 10 eggs before rebounding.  In (c), 100 motiles of age $a_{\textrm{init}} = 0.48$ initialize the population, laying 376 diapause eggs.  In (d), 100 motiles of age $a_{\textrm{init}} = 0.59$ initialize the population, laying 1462 diapause eggs.  From that second cohort onward, the one-year growth factor is well-approximated by $R_0$.}
    \label{fig:transient_plot_mean_14_1_amp_15_5_diap_on_08072021}
\end{figure}

This sensitivity of transient behavior in diapause in the low-rank region is illustrated in Fig.~\ref{fig:transient_plot_mean_14_1_amp_15_5_diap_on_08072021}(a-c), which shows plots of total sessile and motile counts over time for three initial conditions using $h = 14.1$, $g = 15.5$ in Eq.~\eqref{eq:temp}.  In this case, $\lambda_1 \approx 9.1$, $|\lambda_2| \approx 10^{-9}$; in each plot, $z_{\textrm{init}} = 240$ (August 28th, 12 AM), and we start with motiles of ages $a_{\textrm{init}} = 0.11$ (a), $0.36$ (b), $0.48$ (c).  Despite all introduced populations starting in the motile stage, their dynamical trajectories are significantly different: the youngest population at the time of introduction fails to reach egg-laying before being killed off by the cold (a), while the oldest lays a large second cohort of 376 eggs (c).  The middle population only manages to lay 10 eggs, so that the second cohort is only 10$\%$ of the size of the first (b).  It is in this scenario--in which the timing of an arriving population allows for a small but not insignificant amount of egg-laying--that the likelihood of establishment is additionally sensitive to the actual size of the starting population.  If 20 motiles arrive but only produce two eggs, establishment is quite unlikely; however, if 200 motiles arrive and produce 20 eggs, that second cohort has a much higher chance of reproducing successfully itself.  Although transient behavior is sensitive to initial conditions, simulations tend to indicate that if a population can, in fact, establish long term, it tends to grow by a factor of $R_0$ between the second and third cohorts and thereafter.  For example, consider Fig.~\ref{fig:transient_plot_mean_14_1_amp_15_5_diap_on_08072021}(d), which shows the four-year population trajectory of 100 initial individuals of motile age $a_{\textrm{init}} = 0.59$ that arrived on day $z_{\textrm{init}} = 240$.  The second cohort (children of the initial population) consists of 1,462 eggs during diapause, while the third cohort consists of 13,483 diapause eggs, a growth factor of 9.2, which approximates $\lambda_1$.  (The same annual growth factor is repeated thereafter.)  Populations adjust to a new phenology almost immediately after the first year/cohort, in keeping with behavior typically induced by a low-rank operator.

\begin{figure}
    \centering
    \includegraphics[width=\linewidth]{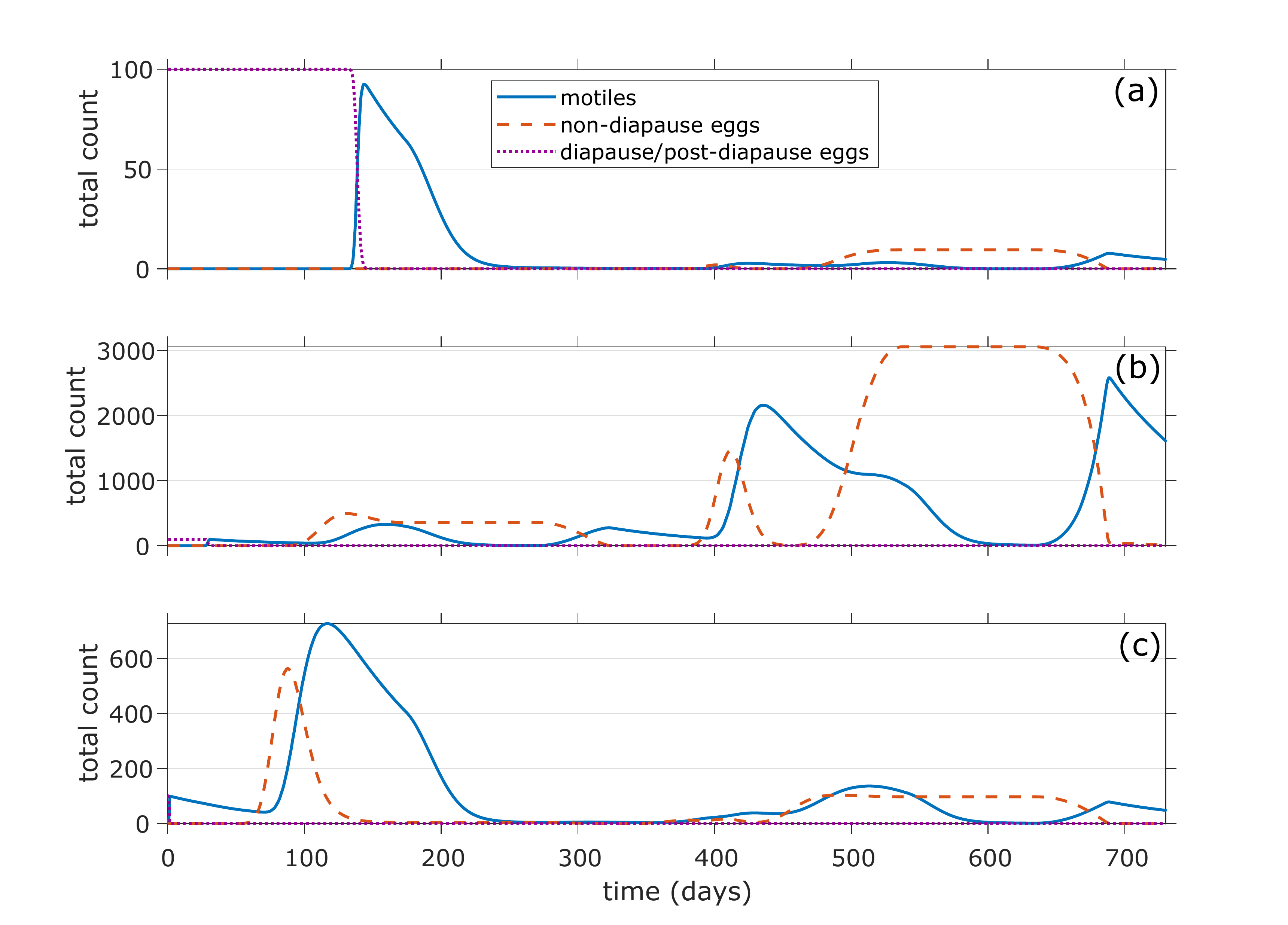}
    \caption{\textbf{Transient Behavior in the Non-diapause High-mean/Low-amplitude Region.} $T(t)$ (Eq.~\eqref{eq:temp}) is parameterized by $h = 17.9$, $g = 11.3$, resulting in $\lambda_1 \approx 5.9$, $\lambda_{2,3} = 0.5 \pm 0.2i$.  In each case, the simulation begins on day of year 165 (June 14th, 12 AM).  In (a), 100 diapause eggs of age $a_{\textrm{init}} = 0.69$ initalize the population, which dies before reaching egg-laying.  In (b), 100 diapause eggs of age $a_{\textrm{init}} = 0.95$ initialize the population, managing to lay a significant second cohort to ensure survival.  In (c), 100 post-diapause eggs of age $a_{\textrm{init}} = 0.95$ initialize the population.  Although it manages to lay a substantial second cohort in the fall, the second cohort dies off in the winter.}
    \label{fig:transient_plot_mean_17_9_amp_11_3_diap_off_08082021}
\end{figure}

The sensitivity of transient behavior to initial conditions in the high-mean/low-amplitude region under the non-diapause model is shown in Fig.~\ref{fig:transient_plot_mean_17_9_amp_11_3_diap_off_08082021}.  Here, $h = 17.9$, $g = 11.3$, with $\lambda_1 = 5.9$, $\lambda_{2,3} = 0.5 \pm 0.2 i$; the four-year sessile and motile counts are plotted for $z_{\textrm{init}} = 165$ (June 14, 12 AM), with individuals starting in diapause ($a_{\textrm{init}} = 0.63$ (a), $0.95$ (b)), and in post-diapause ($a_{\textrm{init}} = 0.95$ (c)).  The younger diapause egg cohort fails to reach egg-laying before being killed by the cold (a), while the older diapause egg cohort lays a substantial second cohort, with strong growth continuing each year thereafter (b).  Due to the timing of the post-diapause cohort's arrival, it lays a substantial second cohort of non-diapause eggs in the fall, but nearly all are killed off in the winter (c).  

\begin{figure}
    \centering
    \includegraphics[width=\linewidth]{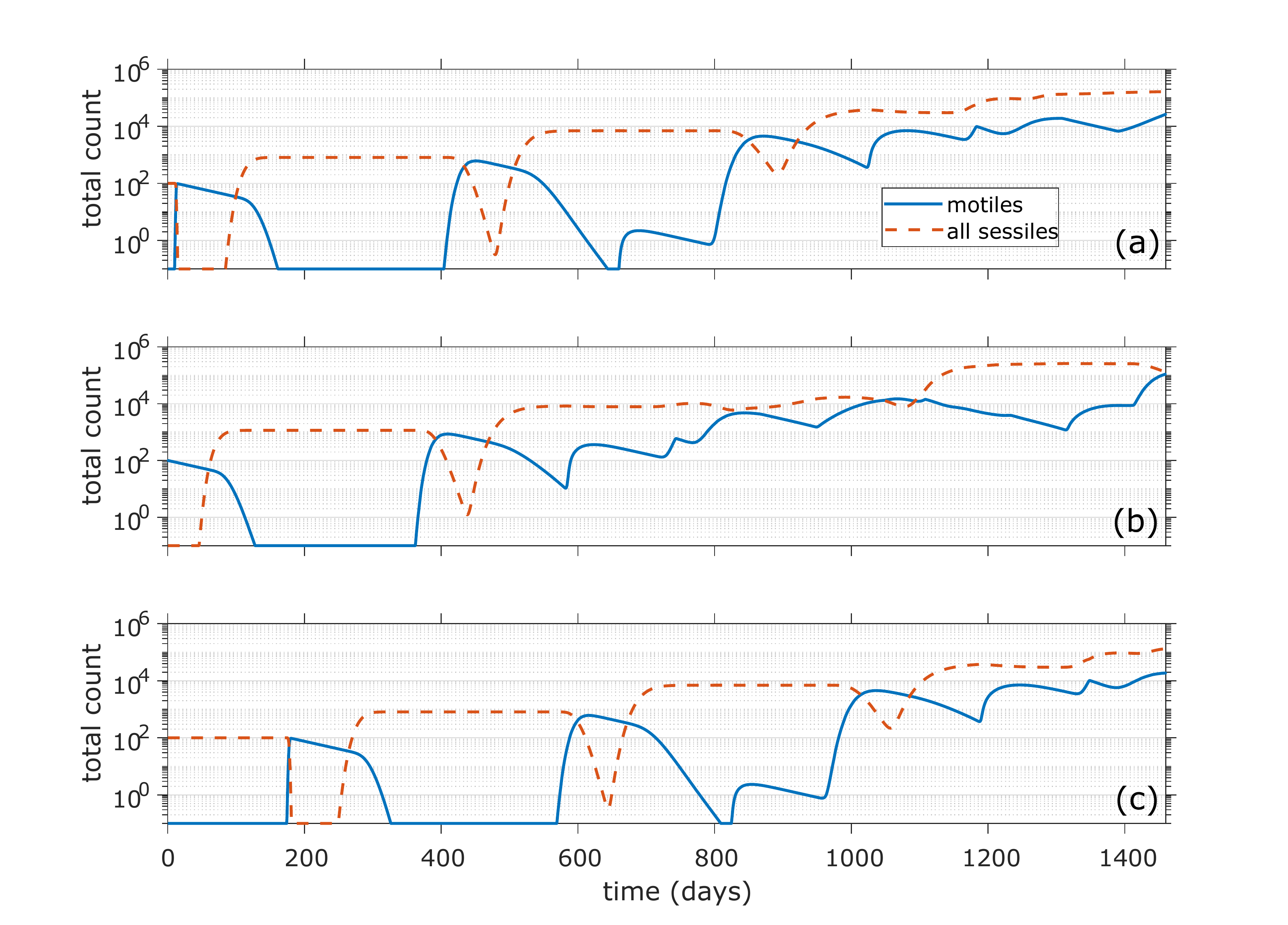}
    \caption{\textbf{Transient Behavior in the Diapause High-mean/Low-amplitude Region.} $T(t)$ (Eq.~\eqref{eq:temp}) is parameterized by $h = 20.9$, $g = 4.8$, resulting in $\lambda_1 \approx 6.7$, $\lambda_{2,3} = -0.5 \pm 5.4i$.  In (a), the population is initialized by 100 non-diapause eggs of age $a_{\textrm{init}} = 0.5$ on day of year 100 (April 10th, 12 AM).  In (b), the population is initialized by 100 second instars of motile age $a_{\textrm{init}} = 0.19$ on day of year 175 (June 12th, 12 AM).  In (c), the population is initialized by 100 diapause eggs of age $a_{\textrm{init}} = 0.49$ on day of year 300 (October 27th, 12 AM).  In all cases, the population experiences strong growth in the transient regime due to warm temperatures and multivoltinism.}
    \label{fig:transient_plot_mean_20_9_amp_4_8_diap_on_08092021}
\end{figure}

Simulations indicate that in the high-mean/low-amplitude region under the diapause model, the presence of several eigenvalues of magnitude larger than 1 generally supports the population through the transient regime, unless $\lambda_1$ is close to 1 itself.  In Fig.~\ref{fig:transient_plot_mean_20_9_amp_4_8_diap_on_08092021}, sessile and motile trajectories are plotted under the diapause model for the temperature profile with $h = 20.9$, $g = 4.8$; in this case, $\lambda_1 = 6.7$, $\lambda_{2,3} = -0.5 \pm 5.4 i$, and we consider several initial conditions that roughly correspond to the phenology in the Pennsylvania establishment area.  In (a), we take $a_{\textrm{init}} = 0.5$ (non-diapause eggs) and $z_{\textrm{init}} = 100$ (April 10th, 12 AM).  In (b), we take $a_{\textrm{init}} = 0.19$ (second instars) and $z_{\textrm{init}} = 175$ (June 24th, 12 AM).  In (c), we take $a_{\textrm{init}} = 0.49$ (diapause eggs) and $z_{\textrm{init}} = 300$ (October 27th, 12 AM).  Transient behavior consists of strong growth from the very beginning for diverse initial conditions, and population establishment does not appear threatened by the period of adjustment to a new phenology.  In each of (a-c), motiles become a constant presence after $\sim 800$ days, the timing of which coincides with the first appearance of a significant population of non-diapause eggs.  In the first two years of each of these simulations, nearly all motiles originate from diapause eggs.  Once the population's phenology shifts enough to allow for non-diapause egg-laying, the population experiences multivoltinism through the non-diapause pathway that supports a constant presence of motiles throughout the year and strong population stability.

Although the set of examples provided here is not comprehensive, it serves to illustrate the potential sensitivity of solutions to transient behavior:~the assessment of establishment risk requires consideration of arrival conditions (timing, developmental age, and initial population size) in addition to $R_0$.  These examples suggest that the presence of multiple eigenvalues larger than 1 in magnitude may provide the growth modes necessary to support transient population survival.  This eigenvalue configuration is prevalent for many high-mean/low-amplitude temperature profiles with and without diapause, underscoring the strong establishment threat in these climates that present both high $R_0$ values and transient growth.

\section{Discussion and Conclusions}
\label{sec:summary}

In this paper, we developed a framework for assessing pest establishment risk in individual locations as a function of the seasonal temperature profile.  To do this, we adapted the existing stage-structured, age-structured PDE system for modeling pest population dynamics to incorporate diapause, a crucial developmental stage for many insects.  We developed generic functional forms of coefficient functions representing the temperature-dependent rates of development, reproduction, and mortality that drive population dynamics, and calibrated them to spotted lanternfly data.  

When our model equations are solved with finite volume methods, the numerical solution experiences artificial diffusion that creates a strong sensitivity to the parameters of the domain discretization.  This is of particular concern with the introduction of diapause into the model; as diapause acts to synchronize the developmental ages of egg populations laid across a season, age distributions with small support tend to arise with seasonal diapause completion.  Numerical diffusion can cause significant spread in such a distribution function, leading to inaccurate egg-laying rates.  To combat this, we designed a novel moving mesh method for simple advection equations, which was used within an operator splitting approach to compute the solution of the model equations.  This new method resulted in solutions that were stable under changes to the discretization parameters, and converged at even coarse discretizations, allowing for efficient computation of solutions.

\begin{figure}
    \centering
    \includegraphics[width=0.8\linewidth]{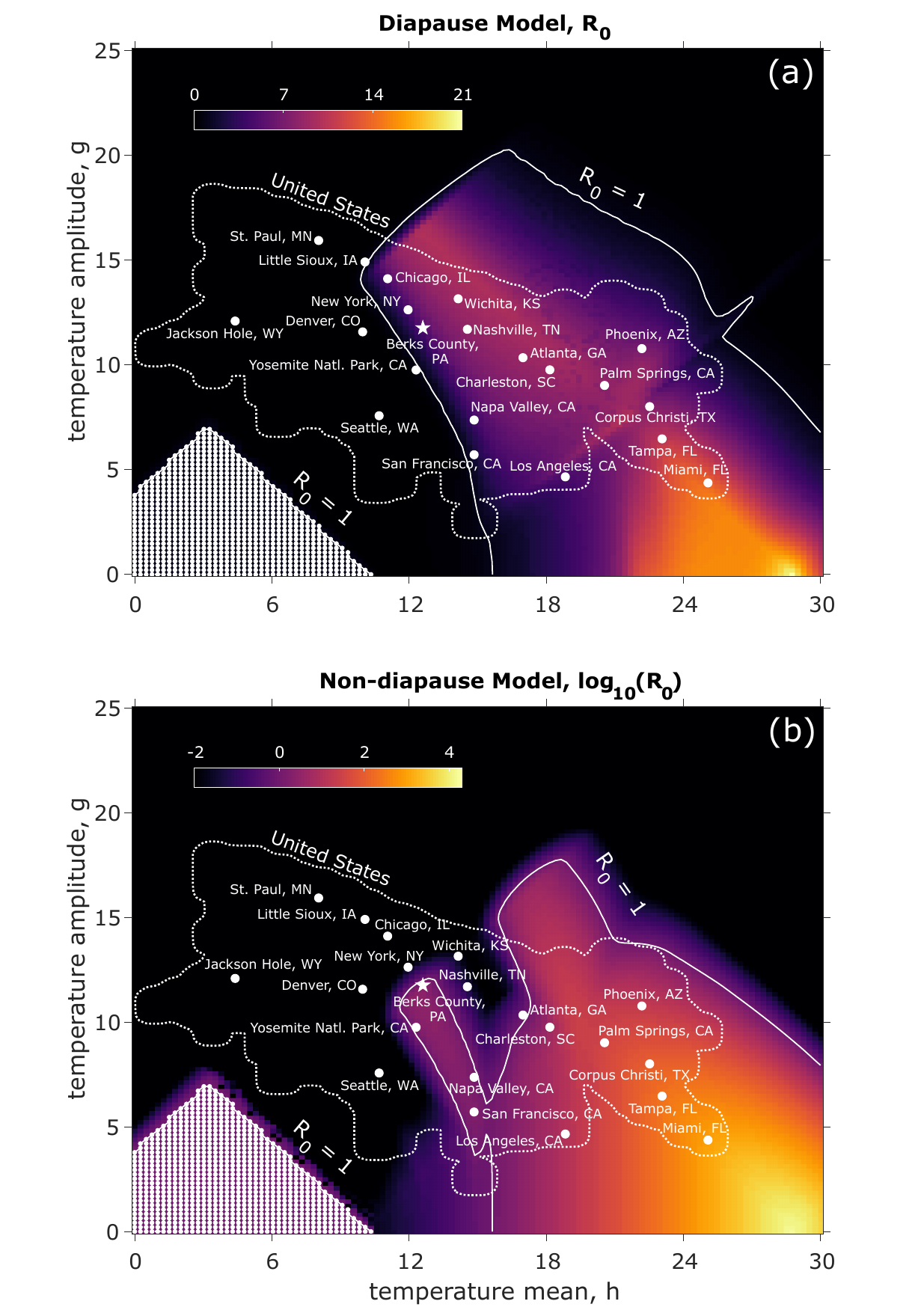}
    \caption{\textbf{$R_0$ in Select United States Locations.}
    (a) and (b) are Figs.~\ref{fig:parametersweepdiapause}(a) and (b), respectively, with locations of interest, as well as the approximate region corresponding to US temperatures (interior of dotted white curve).  Establishment potential is strong in the Southeast and Southwest through both diapause and non-diapause pathways, while only diapausing populations would survive in the Northeast and parts of the South and Midwest.  The Northern Midwest and Pacific Northwest are generally inhospitable to population growth.  This figure was created in MATLAB using the inferno colormap (https://www.mathworks.com/matlabcentral/fileexchange/51986-perceptually-uniform-colormaps)}
    \label{fig:us_cities_r0}
\end{figure}

To assess long-term establishment risk, we derived an approximation to the one-year reproductive number, $R_0$, from the dominant eigenvalue of the one-year numerical solution operator.  To explore the sensitivity of $R_0$ to a broad range of temperature profiles, we modeled temperature with a sinusoidal function and conducted a parameter sweep with respect to mean and amplitude, fixing the phase shift.  Such an analysis provides a complete and concise description of predicted asymptotic establishment patterns, and the $R_0$ value in a real spatial location can be easily assessed by consulting eigenvalue information for the sinusoidal function that best fits average temperature data measured in that place.  In Fig.~\ref{fig:us_cities_r0}, the locations of selected U.S. cities and places of interest in the mean-amplitude parameter domain are superimposed on plots of $R_0$ ($\lambda_1$ from Fig.~\ref{fig:parametersweepdiapause}).  Of particular note is the concentration of several agriculturally significant regions in the moderate-mean/moderate-amplitude part of the parameter domain (for instance, the farming regions of Wichita, KS, and Nashville, TN, and the wine region of Napa Valley, CA).  Diapausing populations have strong potential for establishment in these areas (a), while non-diapause populations have no chance of survival (b).  Several urban areas (New York, NY and Chicago, IL) also present establishment potential for diapausing populations and none for populations incapable of diapause.  The Rocky Mountain areas (e.g. Denver, CO and Jackson Hole, WY) and the northern Midwest (St.~Paul, MN and Little Sioux, IA) tend to possess weak or entirely absent establishment potential for both diapause and non-diapause populations, while the Southeast (Tampa and Miami, FL) and Southwest (Los Angeles and Palm Springs, CA and Phoenix, AZ) have strong establishment potential through both developmental pathways.  In some locations, non-diapause populations can establish, while diapause populations likely cannot (e.g. Yosemite National Park).

In addition to using $\lambda_1$ as a proxy for $R_0$, we consulted the second dominant eigenvalue, $\lambda_2$, to assess population survival and growth in the transient regime.  We found that, under the diapause model, moderate-mean/moderate-amplitude temperature profiles correspond to a low-rank structure in the eigenvalue spectrum, while high-mean/low-amplitude profiles present large $R_0$ values accompanied by several subsequent eigenvalues larger than 1 in magnitude.  Under the non-diapause model, the low-rank region disappears, and $\lambda_1, \lambda_2$ are much larger in magnitude in the high-mean/low-amplitude region than they were in the diapause case.  When $\lambda_1 > 1$ and $|\lambda_2| < 1$, simulations suggest that the single growth mode will not always suffice to support the population through the initial transient regime.  This is the case in the low-rank region in diapause, and on the interior boundaries of the high-mean/low-amplitude region in non-diapause.  When $\lambda_1, |\lambda_2| \gg 1$, on the other hand, populations seem to grow robustly in the transient regime.  This eigenvalue configuration occurs in the high-mean/low-amplitude region of the diapause model, and the interior of that region in the non-diapause model.  Here, the population becomes multivoltine through the non-diapause developmental pathway, with catastrophically high predicted $R_0$ values and strong growth in the transient period.

In addition to informing our understanding of establishment likelihood, these results also offer insight into control strategies.  For moderate-mean/moderate-amplitude temperature profiles in which diapause appears to be necessary for establishment, the first several years following arrival provide a period of potential vulnerability that can be exploited with targeted control measures.  An arriving population may partially collapse in the early stage, and control measures can be timed and targeted to act on the struggling population before it rebounds.  In the high-mean/low-amplitude region, populations grow at an overwhelming rate and control measures are not likely to be effective.  Instead, efforts in these regions should be focused on preventing and interrupting transport of individuals that could found a new population.

It is important to reiterate that the model currently focuses on the effects of temperature on population dynamics, omitting other climatic factors such as soil moisture and evapotranspiration that also affect pest establishment.  Indeed, certain regions, such as the desert climates of the southwestern United States, are likely too dry to host a population despite warm temperatures that drive fast development.  Although the model does incorporate a basal death rate in all four life stages, it does not account for senescence linked to significant developmental delays caused by a mismatch in the phenology of the arriving population and the phenology in its new location.  For instance, in our model, diapausing eggs that arrive in a temperate climate in early spring would likely stay in diapause until the following winter, as they arrived right when temperatures increased and diapause advection slowed.  In reality, those eggs would likely die before exiting diapause.  As further data is collected about the effects of senescence and other climatic factors on population dynamics, the model will be expanded to create an increasingly accurate picture of establishment potential for the spotted lanternfly and other pests of interest.

\section{Acknowledgements}

The authors thank Dennis Calvin and Melody Keena for insightful conversations regarding the calibration of our model to field and experimental data, and Ben Hayes for a helpful conversation on mathematical content.  We thank Victoria Ramirez for her support, and Alexis de la Cotte, Jocelyn Behm, Nad\`{e}ge Belouard, and the Integrative Ecology Lab at Temple University for helpful feedback on this research.  This research includes calculations carried out on HPC resources supported in part by the National Science Foundation through major research instrumentation grant number 1625061 and by the US Army Research Laboratory under contract number W911NF-16-2-0189.   This  work  was  funded  by  the United  States  Department  of  Agriculture  Animal  and  Plant  Health  Inspection  Service  Plant  Protection  and  Quarantine  under  Cooperative   Agreements AP19PPQS$\&$T00C251 and AP20PPQS$\&$T00C136; the United States Department  of  Agriculture  National  Institute  of  Food  and  Agriculture  Specialty  Crop  Research  Initiative  Coordinated  Agricultural  Project  Award  2019-51181-30014; and the Pennsylvania Department of Agriculture under agreements 44176768, 44187342, and C9400000036.

\bibliographystyle{abbrv}
\bibliography{main_bib}

\begin{thebibliography}{10}

\bibitem{NYSIPM2021}
{New York State Integrated Pest Management, Spotted Lanternfly}.
\newblock
  \url{https://nysipm.cornell.edu/environment/invasive-species-exotic-pests/spotted-lanternfly/}.
\newblock Accessed 17 November 2021.

\bibitem{BARRINGER2020}
L.~Barringer and C.~M. Ciafr\'{e}.
\newblock Worldwide {F}eeding {H}ost {P}lants of {S}potted {L}anternfly, {W}ith
  {S}ignificant {A}dditions {F}rom {N}orth {A}merica.
\newblock {\em Environ. Entomol.}, (49):999--1011, 2020.
\newblock \url{https://doi.org/10.1093/ee/nvaa093}.

\bibitem{BARRINGER2015}
L.~E. Barringer, L.~R. Donovall, S.-E. Spichiger, D.~Lynch, and D.~Henry.
\newblock The first {N}ew {W}orld {R}ecord of \textit{{L}ycorma delicatula}
  ({I}nsecta: {H}emiptera: {F}ulgoridae).
\newblock {\em Entomol. News}, (125):20--23, 2015.
\newblock \url{https://doi.org/10.3157/021.125.0105}.

\bibitem{BUFFONI2007}
G.~Buffoni and S.~Pasquali.
\newblock Structured population dynamics:~continuous size and discontinuous
  stage structures.
\newblock {\em J. Math. Biol.}, 54:555--595, 2007.
\newblock \url{https://doi.org/10.1007/S00285-006-0058-2}.

\bibitem{CANELLES2021}
Q.~Canelles, N.~Aquilu\'{e}, P.~M. James, J.~Lawler, and L.~Brontons.
\newblock Global review on interactions between insect pests and other forest
  disturbances.
\newblock {\em Landscape Ecol.}, 36(4):945--972, 2021.
\newblock \url{https://doi.org/10.1007/s10980-021-01209-7}.

\bibitem{CUSHING1994}
J.~M. Cushing.
\newblock The dynamics of hierarchical age-structured populations.
\newblock {\em J. Math. Biol.}, 32:705--729, 1994.
\newblock \url{https://doi.org/10.1007/BF00163023}.

\bibitem{DARA2015}
S.~K. Dara, L.~Barringer, and S.~P. Arthurs.
\newblock Lycorma delicatula ({H}emiptera {F}ulgoridae): {A} {N}ew {I}nvasive
  {P}est in the {U}nited {S}tates.
\newblock {\em J. Integr. Pest Manag.}, 6(1):1--6, 2015.
\newblock \url{https://doi.org/10.1093/JIPM/PMV021}.

\bibitem{EWING2021}
D.~A. Ewing, V.~Blok, and H.~Kettle.
\newblock A process-based, stage-structured model of potato cyst nematode
  population dynamics: Effects of temperature and resistance.
\newblock {\em J. Theoret. Biol.}, 522, 2021.
\newblock \url{https://doi.org/10.1016/j.jtbi.2021.110701}.

\bibitem{FOX1993}
C.~M. Fox.
\newblock The influence of maternal age and mating frequency on egg size and
  offspring performance in \textit{{Callosobruchus maculatus}} ({C}oleoptera:
  {B}ruchidae).
\newblock {\em Oecologia}, 96(1):139--146.
\newblock \url{https://doi.org/10.1007/bf00318042}.

\bibitem{GILIOLI2015}
G.~Gilioli, S.~Pasquali, and E.~Marchesini.
\newblock A modelling framework for pest population dynamics and management: An
  application to the grape berry moth.
\newblock {\em Ecol. Model.}, 320:348--357, 2015.
\newblock \url{https://doi.org/10.1016/j.ecolmodel.2015.10.018}.

\bibitem{GILIOLI2017}
G.~Gilioli, S.~Pasquali, P.~R. Mart\'{i}n, N.~Carlsson, and L.~Mariani.
\newblock A temperature-dependent physiologically based model for the invasive
  apple snail \textit{Pomacea canaliculata}.
\newblock {\em Int. J. Biometerol.}, 61(11):1899--1911, 2017.
\newblock \url{https://doi.org/10.1007/s00484-017-1376-3}.

\bibitem{GILL2017}
H.~K. Gill, G.~Goyal, and G.~Chahil.
\newblock Insect {D}iapause: {A} {R}eview.
\newblock {\em J. Agric. Sci. Technol. A}, 7:454--473, 2017.
\newblock \url{https://doi.org/10.17265/2161-6256/2017.07.002}.

\bibitem{GURTIN1974}
M.~E. Gurtin and R.~C. MacCamy.
\newblock Non-linear age-dependent population dynamics.
\newblock {\em Arch. Ration. Mech. An.}, 54:281--300, 1974.
\newblock \url{https://doi.org/10.1007/BF00250793}.

\bibitem{HE2018}
Z.~He, D.~Ni, and Y.~Liu.
\newblock Theory and approximation of solutions to a harvested hierarchical
  age-structured population model.
\newblock {\em J. Appl. Anal. Comput.}, 8(5):1326--1341, 2018.
\newblock \url{https://doi.org/10.11948/2018.1326}.

\bibitem{IANNELLI1997}
M.~Iannelli, M.-Y. Kim, and E.-J. Park.
\newblock Splitting methods for the numerical approximation of some models of
  age-structured population dynamics and epidemiology.
\newblock {\em Appl. Math. Comput.}, 87(1):69--93, 1997.
\newblock \url{https://doi.org/10.1016/S0096-3003(96)00222-6}.

\bibitem{IANNELLI2017}
M.~Iannelli and F.~Milner.
\newblock {\em {The Basic Approach to Age-structured Population Dynamics}}.
\newblock Springer, 2017.

\bibitem{KAKUMANI2018}
B.~K. Kakumani and S.~K. Tumuluri.
\newblock A numerical scheme to the {McKendrick-von Foerster} equation with
  diffusion in age.
\newblock {\em Numer. Methods Partial Differential Equations}, 00:1--16, 2018.
\newblock \url{https://doi.org/10.1002/num.22280}.

\bibitem{KEENA2021}
M.~A. Keena and A.~L. Nielsen.
\newblock Comparison of the {H}atch of {N}ewly-{L}aid \textit{{L}ycorma
  {D}elicatula} ({H}emiptera: {F}ulgoridae) {E}ggs from the {U}nited {S}tates
  {A}fter {E}xposure to {D}ifferent {T}emperatures and {D}urations of {L}ow
  {T}emperature.
\newblock {\em Environ. Entomol.}, 50(2):410--417, 2021.
\newblock \url{https://doi.org/10.1093/ee/nvaa177}.

\bibitem{KEYFITZ1997}
B.~Keyfitz and N.~Keyfitz.
\newblock The {M}c{K}endrick partial differential equation and its uses in
  epidemiology and population study.
\newblock {\em Math. Comput. Model.}, 26(6):1--9, 1997.
\newblock \url{https://doi.org/10.1016/S0895-7177(97)00165-9}.

\bibitem{KOSTAL2006}
V.~Ko\v{s}t\'{a}l.
\newblock Eco-physiological phases of insect diapause.
\newblock {\em J. Insect Physiol.}, 52(2):113--127, 2006.
\newblock \url{https://doi.org/10.1016/j.jinsphys.2005.09.008}.

\bibitem{KREITMAN2021}
D.~Kreitman, M.~A. Keena, A.~L. Nielsen, G.~Hamilton, and C.~Brent.
\newblock Effects of {T}emperature on {D}evelopment and {S}urvival of
  \textit{Lycorma delicatula} ({H}emiptera: {F}ulgoridae).
\newblock {\em Environ. Entomol.}, 50(1):183--191, 2021.
\newblock \url{https://doi.org/10.1093/ee/nvaa155}.

\bibitem{LAMBRECHTS2011}
L.~Lambrechts, K.~P. Paaijmans, T.~Fansiri, L.~B. Carrington, L.~D. Kramer,
  M.~B. Thomas, and T.~W. Scott.
\newblock Impact of daily temperature fluctuations on dengue virus transmission
  by \textit{{Aedes aegypti}}.
\newblock {\em Proc. Nat. Acad. Sci. USA}, 108(18):7460--7465, 2011.
\newblock \url{https://doi.org/10.1073/pnas.1101377108}.

\bibitem{LEE2019}
D.-H. Lee, Y.-L. Park, and T.~C. Leskey.
\newblock A review of biology and management of {L}ycorma delicatula
  (hemiptera: Fulgoridae), an emerging global invasive species.
\newblock {\em J. Asia-Pac. Entomol.}, 22:589--596, 2019.
\newblock \url{https://doi.org/10.1016/j.aspen.2019.03.004}.

\bibitem{LIU2019}
H.~Liu.
\newblock {Oviposition Substrate Selection, Egg Mass Characteristics, Host
  Preference, and Life History of the Spotted Lanternfly (Hemiptera:
  Fulgoridae) in North America}.
\newblock {\em Environ. Entomol.}, 48:1452--1468, 2019.
\newblock \url{https://doi.org/10.1093/ee/nvz123}.

\bibitem{LIU2017}
K.~Liu, Y.~Lou, and J.~Wu.
\newblock Analysis of an age-structured model for tick populations subject to
  seasonal effects.
\newblock {\em J. Differential Equations}, 263:2078--2112, 2017.
\newblock \url{https://doi.org/10.1016/j.jde.2017.03.038}.

\bibitem{LIU2002}
S.~Liu, L.~Chen, and Z.~Liu.
\newblock Extinction and permanence in nonautonomous competitive system with
  age structure.
\newblock {\em J. Math. Anal. Appl.}, 274:667--684, 2002.
\newblock \url{https://doi.org/10.1016/S0022-247X(02)00329-3}.

\bibitem{LOVETT2016}
G.~M. Lovett, M.~Weiss, A.~M. Liebhold, T.~P. Holmes, B.~Leung, K.~F. Lambert,
  D.~A. Orwig, F.~T. Campbell, J.~Rosenthal, D.~G. McCullough, R.~Wildova,
  M.~P. Ayres, C.~D. Canham, D.~R. Foster, S.~D. LaDeau, and T.~Weldy.
\newblock Nonnative forest insects and pathogens in the united states:
  {I}mpacts and policy options.
\newblock {\em Ecol. Appl.}, 26(5):1437--1455, 2016.
\newblock \url{https://doi.org/10.1890/15-1176}.

\bibitem{MOUSSEAU1991}
T.~A. Mousseau and H.~Dingle.
\newblock Maternal effects in insect life histories.
\newblock {\em Annu. Rev. Entomol.}, (36):511--534, 1991.
\newblock \url{https://doi.org/10.1146/annurev.en.36.010191.002455}.

\bibitem{NOWAK2000}
M.~A. Nowak and R.~M. May.
\newblock {\em Virus dynamics: mathematical principles of immunology and
  virology}.
\newblock Oxford University Press, 2000.

\bibitem{PARK2015}
M.~Park.
\newblock {\em Overwintering ecology and population genetics of {L}ycorma
  delicatula (Hemiptera: Fulgoridae) in Korea}.
\newblock PhD thesis, Seoul National University, 2015.

\bibitem{PASQUALI2019}
S.~Pasquali, C.~Soresina, and G.~Gilioli.
\newblock The effects of fecundity, mortality, and distribution of the initial
  condition in phenological models.
\newblock {\em Ecol. Model.}, 402:45--58, 2019.
\newblock \url{https://doi.org/10.1016/j.ecolmodel.2019.03.019}.

\bibitem{PURESWARAN2018}
D.~S. Pureswaran, A.~Roques, and A.~Battisti.
\newblock Forest {I}nsects and {C}limate {C}hange.
\newblock {\em Curr. For. Rep.}, 4(2):35--50, 2018.
\newblock \url{https://link.springer.com/article/10.1007/s40725-018-0075-6}.

\bibitem{SAUNDERS2004}
D.~Saunders, R.~Lewis, and G.~Warman.
\newblock Photoperiodic induction of diapause: opening the black box.
\newblock {\em Physiol. Entomol.}, (29):1--15, 2004.
\newblock \url{https://doi.org/10.1111/j.1365-3032.2004.0369.x}.

\bibitem{SAUNDERS1981}
D.~S. Saunders.
\newblock Insect photoperiodism — the clock and the counter: a review.
\newblock {\em Physiol. Entomol.}, 6:99--116, 1981.
\newblock \url{https://doi.org/10.1111/j.1365-3032.1981.tb00264.x}.

\bibitem{SHARPE1911}
F.~Sharpe and A.~Lotka.
\newblock A problem in age-distribution.
\newblock {\em Lond. Edinb. Dubl. Phil. Mag.}, 21(124):435--438, 1911.
\newblock \url{https://doi.org/10.1080/14786440408637050}.

\bibitem{SHIM2015}
J.-K. Shim and K.-Y. Lee.
\newblock Molecular characterization of heat shock protein 70 cognate c{DNA}
  and its upregulation after diapause termination in \textit{{L}ycorma
  delicatula} eggs.
\newblock {\em J. Asia-Pac. Entomol.}, 18(4):709--714.
\newblock \url{https://doi.org/10.1016/j.aspen.2015.08.005}.

\bibitem{SLADONJA2015}
B.~Sladonja, M.~Su\v{s}ek, and J.~Guillermic.
\newblock Review on {I}nvasive {T}ree of {H}eaven (\textit{Ailanthis
  altissima}~({M}ill.) {S}wingle) {C}onflicting {V}alues: {A}ssessment of {I}ts
  {E}cosystem {S}ervices and {P}otential {B}iological {T}hreat.
\newblock {\em Environ. Manage.}, (56):1009--1034, 2015.
\newblock \url{https://doi.org/10.1007/s00267-015-0546-5}.

\bibitem{SMYERS2021}
E.~C. Smyers, J.~M. Urban, A.~C. Dechaine, D.~G. Pfeiffer, S.~R. Crawford, and
  D.~D. Calvin.
\newblock Spatio-{T}emporal {M}odel for {P}redicting {S}pring {H}atch of the
  {S}potted {L}anternfly ({H}emiptera: {F}ulgoridae).
\newblock {\em Environ. Entomol.}, 50(1):126--137, 2021.
\newblock \url{https://doi.org/10.1093/ee/nvaa129}.

\bibitem{STEINER2014}
U.~K. Steiner, S.~Tuljapurkar, and T.~Coulson.
\newblock {Generation Time, Net Reproductive Rate, and Growth in Stage-age
  Structured Populations}.
\newblock {\em Am. Nat.}, 183(6):771--783, 2014.
\newblock \url{https://doi.org/10.1086/675894}.

\bibitem{TAUBER1976}
M.~J. Tauber and C.~A. Tauber.
\newblock {Insect Seasonality: Diapause Maintenance, Termination, and
  Postdiapause Development}.
\newblock {\em Annu. Rev. Entomol.}, 21:81--107, 1976.
\newblock \url{https://doi.org/10.1146/annurev.en.21.010176.000501}.

\bibitem{THOMAS2012}
S.~M. Thomas, U.~Obermayr, D.~Fischer, J.~Kreyling, and C.~Beierkuhnlein.
\newblock Low temperature threshold for egg survival of a post-diapause and
  non-diapause {E}uropean aedine strain, \textit{{A}edes albopictus}
  ({D}iptera: {C}ulicidea).
\newblock {\em Parasite Vector.}, 5(1):100, 2012.
\newblock \url{https://doi.org/10.1186/1756-3305-5-100}.

\bibitem{URBAN2019}
J.~M. Urban.
\newblock Perspective: shedding light on spotted lanternfly impacts in the
  {USA}.
\newblock {\em Pest Manag. Sci.}, 76(1):10--17, 2019.
\newblock \url{https://doi.org/10.1002/ps.5619}.

\bibitem{WAGNER1984}
T.~L. Wagner, H.-I. Wu, P.~J. Sharpe, R.~M. Schoolfield, and R.~N. Coulson.
\newblock Modeling {I}nsect {D}evelopment {R}ates: a {L}iterature {R}eview and
  {A}pplication of a {B}iophysical {M}odel.
\newblock {\em Ann. Entomol. Soc. Am.}, 77(2):208--220, 1984.
\newblock \url{https://doi.org/10.1093/aesa/77.2.208}.

\bibitem{WANG2016}
X.~Wang, S.~Tang, and R.~A. Cheke.
\newblock A stage structured mosquito model incorporating effects of
  precipitation and daily temperature fluctuations.
\newblock {\em J. Theoret. Biol.}, 411:27--36, 2016.
\newblock \url{https://doi.org/10.1016/j.jtbi.2016.09.015}.

\bibitem{WOLFIN2019}
M.~S. Wolfin, M.~Binyameen, Y.~Wang, J.~M. Urban, D.~C. Roberts, and T.~C.
  Baker.
\newblock Flight {D}ispersal {C}apabilites of {F}emale {S}potted {L}anternflies
  (\textit{{L}ycorma delicatula}) {R}elated to {S}ize and {M}ating {S}tatus.
\newblock {\em J. Insect Behav.}, 32(3):188--200, 2019.
\newblock \url{https://link.springer.com/article/10.1007/s10905-019-09724-x}.

\end{thebibliography}

\end{document}